\documentclass{article}

\usepackage{PRIMEarxiv}

\usepackage[utf8]{inputenc} % allow utf-8 input
\usepackage[T1]{fontenc}    % use 8-bit T1 fonts
\usepackage{hyperref}       % hyperlinks
\usepackage{url}            % simple URL typesetting
\usepackage{booktabs}       % professional-quality tables
\usepackage{amsfonts}       % blackboard math symbols
\usepackage{nicefrac}       % compact symbols for 1/2, etc.
\usepackage{microtype}      % microtypography
\usepackage{lipsum}
\usepackage{fancyhdr}       % header
\usepackage{graphicx}       % graphics
\usepackage{multirow} 
\usepackage{authblk}
\usepackage{color}
\usepackage{hyperref}
\hypersetup{
    colorlinks=true,
    linkcolor=blue,
    filecolor=blue,      
    urlcolor=blue,
    citecolor=cyan,
}

\makeatletter
\newcommand{\ssymbol}[1]{^{\@fnsymbol{#1}}}
\makeatother
\graphicspath{{media/}}     % organize your images and other figures under media/ folder

\title{AccidentGPT: Accident Analysis and Prevention from V2X Environmental Perception with Multi-modal Large Model}

\begin{document}

\author[a]{$Lening~Wang$}
\author[a, d]{ $Yilong~Ren ^{*}$}
\author[a]{$Han~Jiang$}
\author[b]{$Pinlong~Cai$}
\author[b]{$Daocheng~Fu$}
\author[c]{$Tianqi~Wang$}
\author[a]{$Zhiyong~Cui ^{*}$}
\author[a, d]{ $ Haiyang~Yu ^{*}$ }
\author[e]{ $ Xuesong~Wang $ }
\author[f]{ $ Hanchu~Zhou $ }
\author[f]{ $ Helai~Huang $ }
\author[g]{ $ Yinhai~Wang $ }

\affil[a]{State Key Lab of Intelligent Transportation System, Beihang University, Beijing, China, 100191\authorcr {\{leningwang; yilongren; buaajh; zhiyongc; hyyu\}} @buaa.edu.cn}
\affil[b]{Shanghai Artificial Intelligence Laboratory, Shanghai, China, 200232 \authorcr {\{caipinlong; fudaocheng\}}@pjlab.org.cn }
\affil[c]{The University of Hong Kong, Hong Kong, China, 999077  \authorcr wangtq@connect.hku.hk}
\affil[d]{Zhongguancun Laboratory, Beijing, China, 100094}
\affil[e]{Tongji University, Shanghai, China, 201804  \authorcr wangxs@tongji.edu.cn }
\affil[f]{Central South University, Changsha, China, 410075  \authorcr {\{hanchuzhou; huanghelai\} }@csu.edu.cn }
\affil[g]{University of Washington, Seattle, United States, 98195-2700 \authorcr yinhai@uw.edu}

\affil[*]{Corresponding author  }

\maketitle

\begin{abstract}

Traffic accidents, being a significant contributor to both human casualties and property damage, have long been a focal point of research for many scholars in the field of traffic safety. However, previous studies, whether focusing on static environmental assessments or dynamic driving analyses, as well as pre-accident predictions or post-accident rule analyses, have typically been conducted in isolation. There has been a lack of an effective framework for developing a comprehensive understanding and application of traffic safety. To address this gap, this paper introduces AccidentGPT, a comprehensive accident analysis and prevention multi-modal large model. AccidentGPT establishes a multi-modal information interaction framework grounded in multi-sensor perception, thereby enabling a holistic approach to accident analysis and prevention in the field of traffic safety. Specifically, our capabilities can be categorized as follows: for autonomous driving vehicles, we provide comprehensive environmental perception and understanding to control the vehicle and avoid collisions. For human-driven vehicles, we offer proactive long-range safety warnings and blind-spot alerts while also providing safety driving recommendations and behavioral norms through human-machine dialogue and interaction. Additionally, for traffic police and management agencies, our framework supports intelligent and real-time analysis of traffic safety, encompassing pedestrian, vehicles, roads, and the environment through collaborative perception from multiple vehicles and road testing devices. The system is also capable of providing a thorough analysis of accident causes and liability after vehicle collisions. Our framework stands as the first large model to integrate comprehensive scene understanding into traffic safety studies. Project page: \url{https://accidentgpt.github.io}
\end{abstract}

% keywords can be removed

\section{Introduction}

Like the majority of research in traffic safety, the study of accident analysis and prevention, aiming to enhance the safety of the traffic system and reduce casualties and property losses, is a necessary endeavor \cite{who_health_topics, hu2022high, ren2022tbsm}. In previous research, scholars have approached traffic safety from various perspectives, categorizing their studies according to traffic scenarios, events, and degrees of traffic intelligence. These investigations broadly encompass a range of focal areas: analysis of static environments \cite{an2022examining, asadi2022comprehensive}, examination of dynamic object states \cite{wang2021real}, forecasting and analysis of accidents and their causes, evaluation of human driving safety \cite{ali2020impact, wang2022analysis}, and in-depth analysis of comprehensive safety in autonomous driving \cite{ye2021approaching, ryerson2019edge}. This multifaceted approach reflects the complexity and breadth of traffic safety research. For instance, in the realm of static environment analysis, a thorough examination is conducted to understand the impact of traffic safety within the broader context of road and environmental scenarios. In dynamic object state analysis, the emphasis lies on aspects like collision and conflict prevention, the study of driving behaviors, and the modeling of vehicle dynamics. Several researchers focus on accidents, engaging in proactive collision prediction before incidents and in-depth causation analysis afterwards, with the aim to reduce the frequency and severity of traffic accidents. Concurrently, as autonomous driving research rapidly advances, the focus has shifted from traditional manual driving safety towards investigating the safety of human-machine hybrid driving \cite{ansari2022human} transitions and the coexistence of mixed vehicle types. This shift aims to ensure precision in safety and control mechanisms for fully autonomous vehicles.

However, despite a common overarching goal of fostering a safer, more comprehensive traffic safety environment and preventing accidents, past research efforts have often been compartmentalized. This resulted in a fragmentation of efforts, with a notable absence of effective coordination between different tasks and modules. Consequently, there has been a lack of a unified, integrated framework capable of achieving a holistic understanding and practical application of traffic safety principles.

With the swift progression of artificial intelligence, the integration of collaborative perception \cite{wang2023deepaccident,han2023collaborative,hu2022where2comm} and Large Language Model (LLM) \cite{openai2023gpt,wen2023dilu,zhang2023trafficgpt} promises a deeper understanding and enriched interaction in traffic safety. To this end, we introduce AccidentGPT, an advanced multi-modal large model designed for accident analysis and prevention, rooted in Environmental Perception. This framework is meticulously engineered to integrate and interpret a wide array of sensor data, forming a multimodal platform for interaction and comprehension. Its primary objective is to address accident analysis and prevention, smoothing the shift from human-operated driving to intelligent vehicle safety in autonomous driving.

Specifically, AccidentGPT establishes the V2X-perception framework for exhaustive scene perception and comprehension, complemented by the GPT-reasoning framework for multi-modal interaction across various specialized tasks. This leads to a holistic approach in traffic safety research. The Vehicle-to-Everything (V2X) perception framework amalgamates panoramic imagery from multiple vehicles and road-testing devices, enabling 3D object detection, Bird's Eye View (BEV) perspective perception, and the recognition and construction of comprehensive scenarios involving pedestrian, vehicles, roads, and the environment. This approach not only augments perception and planning for autonomous driving as well as achieves an all-encompassing understanding of traffic safety scenes. It aids in real-time assessment of the traffic safety environment, alongside accident prediction and analysis. Moreover, within the GPT-reasoning framework, we draw from the innate cognition of human driving and extensive traffic safety analysis. This structure deeply integrates a LLM, merging cognitive reasoning with perceptual skills. It incorporates various traffic safety tasks into its corpus, enabling profound interactions in natural language reasoning and perception tasks. 

Our contributions can be mainly divided into the following parts:

\begin{itemize}

    \item  Autonomous Vehicle Environmental Perception: We have developed a comprehensive scene perception and prediction system for fully autonomous vehicles. This system uses cameras on multiple vehicles and road-testing devices to create a unified BEV space, enabling 3D object detection, perspective analysis, and motion prediction. Our approach integrates multi-vehicle collaborative perception for enhanced environmental understanding and collision avoidance.

    \item  Safety Enhancements for Human-driven Vehicles: For human-driven vehicles, we offer advanced safety features such as proactive remote safety warnings and blind spot alerts. Our system comprises distinct agents, labeling, and semantic modules that process visual data, including scene, BEV images, and 3D objects detection results. This multi-level modular input structure, supported by a corporate database and priority sampling module, improves the versatility of safety measures for human-driven vehicles.

    \item  Traffic Safety Analysis for Enforcement Agencies: Our framework serves traffic police and management agencies by providing real-time intelligent analysis of traffic safety factors – pedestrian, vehicles, roads, and the environment. Utilizing collaborative perception from multiple vehicles and road testing devices, it delivers comprehensive reports on accident causes and responsibilities. The system's effectiveness is enhanced by a rich pre-training on an efficient traffic safety corpus, ensuring accurate analysis and insights. % insights

\end{itemize}

These points encapsulate the primary contributions of our capabilities, highlighting our advancements in autonomous vehicle technology, safety features for human-driven vehicles, and traffic safety analysis tools for enforcement agencies. In summary, our framework supports all-around traffic safety research and is the first to integrate comprehensive scene understanding into traffic safety research using a large model.

\section{Related Work}
\label{Related Work}

Our research is situated at the intersection of traffic safety analysis and warning tasks, autonomous driving perception and planning tasks, and the application of LLMs in the field of transportation and vehicles. Therefore, we will separately introduce the mainstream research in these three tasks and the points of integration in our research.

% \lipsum[4] See Section \ref{sec:headings}.

\subsection{Traffic safety analysis and accident warning
}

According to the World Health Organization, approximately 1.3 million people die annually due to road traffic collisions \cite{who_health_topics}. Consequently, research on traffic safety remains a focal point in the fields of transportation and vehicle engineering. Particularly with the emergence of autonomous driving research, studies in traffic safety under autonomous or human-machine collaborative driving have become increasingly prominent.

Traffic safety and accident analysis cover a wide range of aspects. Broadly, they can be divided into static environment analysis and dynamic object state inspection in spatial terms, and pre-collision prediction and post-accident analysis in temporal terms. In static environment analysis, scholars have conducted comprehensive assessments and analyses of factors affecting traffic safety, including elements like pedestrian, vehicles, roads, and the environment \cite{obelheiro2020new, asadi2022comprehensive, stevanovic2015multi}. These methods, although subjective, have been shown to reduce vehicle collisions and enhance traffic safety to some extent in redesigned and adjusted road environments, as evidenced by empirical studies. Additionally, in the area of dynamic object safety research, scholars have delved into modeling vehicle conflict and analyzing driver behavior, focusing on aspects like vehicle following relationships \cite{zheng2020analyzing}, driver compliance levels \cite{sharma2021assessing}, and safety inspections in mixed environments at intersections \cite{arvin2020safety}.

Moreover, as accidents mark a causal boundary in traffic safety, pre-collision prediction and post-accident analysis are extremely important. Researchers have used various deep learning techniques, such as graph neural networks, to design traffic accident prediction networks \cite{yu2021deep, lin2021intelligent}, aiming to predict traffic incidents. Some studies \cite{wang2023deepaccident} analyze vehicle trajectories to assess the likelihood of collisions, striving to prevent them. Furthermore, analyzing accident data to identify causes and improve traffic safety is vital, whether through multivariate analysis of road traffic accident severity \cite{wang2019road} or examining correlations between potential risk factors using data mining algorithms \cite{alkheder2020risk}. With the advent and evolution of autonomous vehicles, research exploring the factors influencing the severity of collisions involving autonomous cars \cite {ren2022divergent} is emerging, seeking to assess the different impacts of autonomous and human driving patterns on collision severity for safety evaluation.

However, past research \cite{yu2021deep, wang2019road, alkheder2020risk}, whether based on comprehensive static environment analysis, dynamic vehicular kinematics or dynamics modeling, or through accident prediction and post-event evaluation, has been relatively independent. There lacks a holistic framework encompassing all these aspects, especially considering the advanced state of perception, prediction, and planning in autonomous driving. Proposing a complete end-to-end, comprehensive study of traffic safety is thus imperative.

\subsection{Perception and planning}

The mainstream approach in autonomous driving research has been to detect, recognize, and classify targets in the surrounding environment using sensors carried by vehicles or road-testing devices. This process serves the planning and control of vehicles. In the domain of target detection, research can be categorized into three main directions: single-vehicle intelligent perception, multi-vehicle collaborative perception, and vehicle-road collaborative perception. 

In the realm of single-vehicle intelligent perception, landmark studies such as CenterNet \cite{duan2019centernet} proposed a keypoint detection method, offering an efficient solution to explore visual patterns within individual cropped regions at minimal cost. ZoomNet \cite{xu2020zoomnet} introduced a vehicle-centric camera-based target detection method with an adaptive zoom module, predicting higher-quality disparities from transformed recognition boxes. RTM3D \cite{li2020rtm3d} considered multiple perspective keypoints for 3D bounding boxes in image space, improving target detection rates in terms of speed and efficiency. FCOS3D \cite{wang2021fcos3d} introduced a universal 3D target detection framework, decoupling 2D and 3D attributes from defined 7-DOF 3D target boxes, achieving top performance in visual-only 3D detection in the nuScenes dataset.

Recognizing the limitations of single-vehicle perception, some studies have focused on adopting a multi-vehicle collaborative approach to perception. CoBEVT \cite{xu2022cobevt} leveraged Vehicle-to-Vehicle (V2V) communication to share sensor information among vehicles, significantly improving perception performance and range, thereby collaboratively generating predictions of BEV maps. Coopernaut \cite{Cui_2022_CVPR} similarly constructed a collaborative perception framework using vehicle-to-vehicle communication for vision-based cooperative driving. It incorporated laser radar information, encoding it into a compact point-based representation for transmission between vehicles via real wireless channels. BEV-V2X \cite{10179171} proposed an innovative approach based on V2X communication. It collected local BEV data from all connected and autonomous vehicles in the control area through roadside units or cloud centers, enabling the fusion and prediction of future global BEV occupancy grid maps. V2VFormer++ \cite{yin2023v2vformer} built the first multimodal vehicle-to-vehicle collaborative perception framework, fusing individual camera-lidar representations with Dynamic Channel Fusion (DCF) in BEV space to capture global responses and achieve long-range collaborative perception.

Moreover, as road-testing devices capture higher angles and broader perspectives, covering specific scenes that individual vehicles may miss, some research has concentrated on integrating vehicle-road collaborative perception. To achieve Level 5 autonomous driving, the DAIR-V2X \cite{yu2022dair} dataset emphasized research on vehicle-infrastructure cooperation, releasing a large-scale, multimodal, multi-view vehicle-road collaborative dataset. BEVHeight \cite{yang2023bevheight} proposed a method for fusing vehicle-road collaborative perception, considering the rapid reduction in depth differences between vehicles and the ground as distance increases. It projected the height of road-testing cameras pixel-wise to depth, enhancing accuracy in occluded viewpoints and target detection under pose transformations. FecNet \cite{10223730} fully considered static backgrounds captured by roadside lidar, presenting a feature-enhanced cascaded network. It improved the recognition ability of foreground features by extracting foreground information and fusing it with multi-stage feature maps.

In fact, past research on perception has mainly focused on the three mainstream directions of single-vehicle intelligent perception, multi-vehicle collaborative perception, and vehicle-road collaborative perception. However, research on integrating multi-vehicle collaborative perception with vehicle-road collaborative perception is still rare. If multiple devices with different perspectives and ranges can be further integrated, it will undoubtedly expand the range of perception and improve its accuracy.

	\subsection{LLM in transportation and intelligent vehicles}

In the realm of LLM research, with ChatGPT as a prominent example, a new era in semantic parsing has been ushered in. Leading LLMs are generally based on transformer architectures for natural language processing, often equipped with tens of millions or even billions of parameters. These models are trained on extensive text datasets, enabling them to comprehend natural language and perform a wide range of complex tasks. The latest LLMs now possess dual capabilities in understanding both text and images. TransGPT introduces a versatile common-sense traffic model capable of predicting traffic situations, acting as an intelligent advisory assistant, supporting public transportation services, aiding in traffic planning and design, facilitating traffic safety education, assisting in management tasks, handling traffic accident reporting and analysis, and contributing to autonomous driving assistance systems.

The DiLu framework, as proposed by Wen et al. \cite{wen2023dilu}, combines reasoning and reflection modules, allowing the system to make decisions based on common-sense knowledge and evolve continuously. This infusion of knowledge-driven capabilities into autonomous driving systems, considering how humans drive, enables functionalities such as lane changing. TrafficGPT, as presented by Zhang et al. \cite{zhang2023trafficgpt}, integrates ChatGPT with foundational traffic models, enhancing the model's ability to view, analyze, and process traffic data. This integration provides insightful decision support for managing urban traffic systems. Wang et al. \cite{wang2023chatgpt} design a universal framework embedding LLM into the co-pilot role of vehicles. This framework enables the completion of specific driving tasks based on provided information, aligning with human intentions. 

Previous work on LLM in the fields of traffic and autonomous driving has mainly focused on fine-tuning and training LLMs using traffic data. Alternatively, it involved secondary development of convenience applications based on LLM prompts or planning and control of vehicle movement based on LLM prompt words. However, the outstanding performance of LLM lies in its superior reasoning and judgment capabilities. Therefore, our in-depth integration and interaction of perception and trajectory prediction planning models with LLM aim to combine the perceptual work of computers with LLM research. This approach facilitates diverse tasks under human-machine interaction, including joint traffic situation warnings, accident responsibility allocation, accident causation analysis, and human-machine interaction prompts. This comprehensive integration effectively enhances the significance and value of perceptual work for both autonomous driving and human-machine collaborative driving, thereby advancing research and applications in traffic safety within the intelligent connected vehicle environment.
	
% ``''

%崔志勇:
%首先，AccidentGPT要做的事儿还有目标，感觉描述了，但不是很具体清晰，或者说缺少例子，可以补充点例子（包括图和文字叙述），写得更具体点，有益于理解

%崔志勇:
%方法、module、component部分，我觉得需要有重点、有强调。以module举例，重点强调为什么要加某个module，这个module是用来做什么，怎么做，然后再说里面有啥组件。现在写的有点流水账，还得尽快再磨一磨 

%崔志勇:
%最核心的感觉是，看完methodology，这个框架是干嘛用的，有点模糊，后面appendix里面的示例图需要往前提，需要让读者尽快知道这个AccidentGPT在干嘛，我觉得 第一个图是框架图，第二个图就可以是功能示例图。而且功能的示例图相比于其他一些文章，可以再美化一下

\begin{figure}[t]
  \centering

\includegraphics[width=1\linewidth]{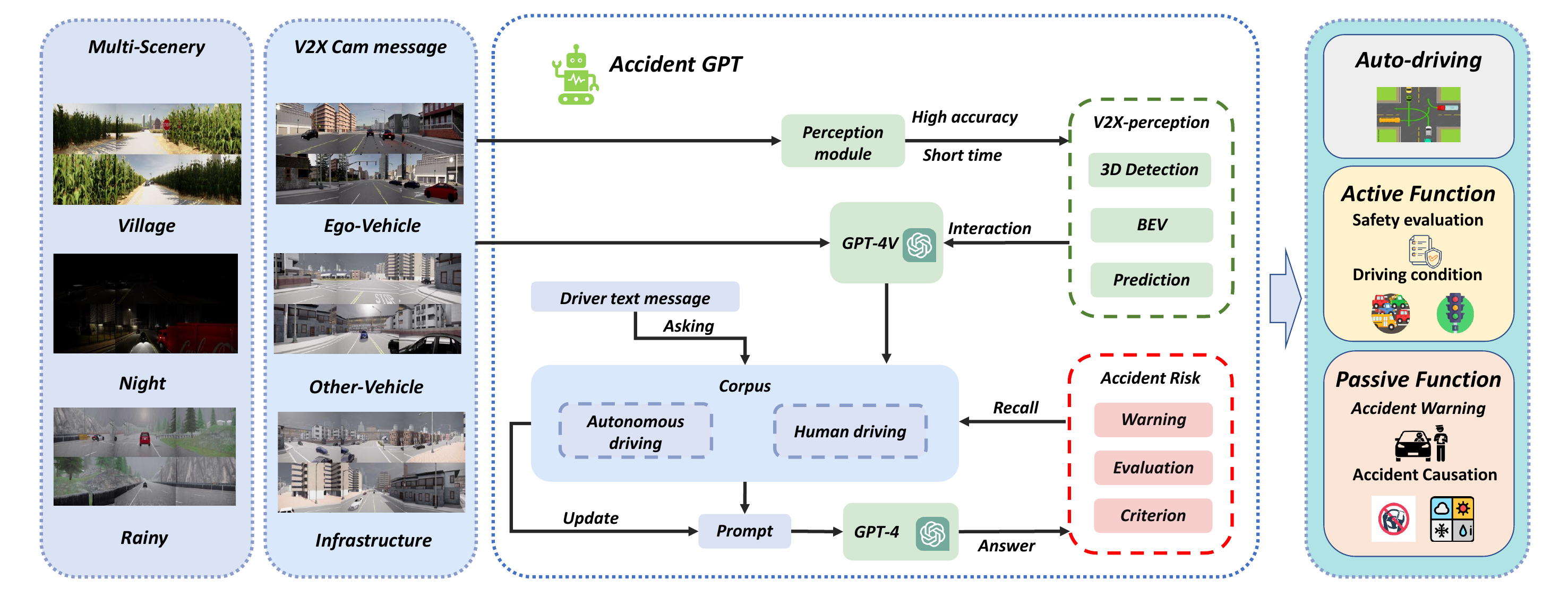}
  \caption{The overall framework of AccidentGPT.}
  \label{fig:fig1}
\end{figure}

\section{Methodology}
\label{Methodology}

\subsection{Overview}
	\label{Overview}

The Overview of AccidentGPT, as depicted in Figures \ref{fig:fig1}, presents a comprehensive structure emphasizing both the mission and the functional components of the system. The framework is designed to enhance traffic safety research, including the human-machine interaction safety dialogue for artificially driven vehicles and traffic safety planning research for relevant departments.

AccidentGPT primarily consists of two main components. The first, V2X-perception, acts as the comprehensive scene perception module within the V2X architecture. It constructs an image extraction and analysis network using panoramic multi-camera information from multiple vehicles and roadside devices. Key elements of this module include the image processing backbone, view-transformer for image pose transformation, fusion-BEV for multi-scene fused perception, and a multi-head module for target detection, BEV perception, and trajectory prediction.

The second core component is the GPT-reasoning module, functioning as the reasoning interaction module under natural language. This includes the standard GPT-4V module, the multi-level priority sampling prompt module under corpus fusion, the active-passive task prompt module, and multiple output modules for accident-related tasks. This configuration aims to facilitate a seamless human-machine collaboration, leveraging machine language perception and human thought-driven knowledge to improve traffic safety.

To better elucidate our approach, we provide precise definitions for the proposed tasks as depicted in Figures \ref{fig:fig5}. For the perception module, the task involves generating 3D object detection boxes, BEV perception, and trajectory prediction, utilizing inputs from multiple devices and perspectives. In terms of natural language reasoning tasks, the focus is on using the input and output information from the perception tasks as queried by the query module. This also includes constructing a learnable and updatable LLM corpus for prompts, enabling proactive prompts through drivers' active inquiries and passive prompts from the model, offering drivers passive cues such as advance notices and warnings for traffic safety events. Additionally, it involves responding to drivers' active inquiries and addressing tasks initiated by traffic authorities, like road traffic safety risk assessments and determining responsibilities in traffic accidents.

This comprehensive framework aims to address various tasks related to traffic and vehicle safety, particularly in the contexts of autonomous driving and human-machine interaction.
 
%The overall framework of AccidentGPT, as depicted in Figure \ref{fig:fig1}, consists primarily of two main components. Firstly, the V2X-perception acts as the comprehensive scene perception module under the V2X architecture. It constructs an image extraction and analysis network for panoramic multi-camera information from multiple vehicles and roadside devices. This includes the image processing backbone, view-transformer for image pose transformation, fusion-BEV for multi-scene fused perception, and the multi-head module for target detection, BEV perception, and trajectory prediction. The second component is GPT-reasoning, serving as the reasoning interaction module under natural language. Key elements within this module include the standard GPT-4V module, the Multi-level Priority Sampling Prompt module under corpus fusion, the Active-Passive Task Prompt module, and multiple output modules for accident-related tasks. This configuration aims to facilitate human-machine collaboration under machine language perception and human thought-driven knowledge. It enhances overall traffic safety research, including tasks related to the human-machine interaction safety dialogue for artificially driven vehicles and traffic safety planning research for relevant departments.

\begin{figure}[h]
  \centering

\includegraphics[width=1\linewidth]{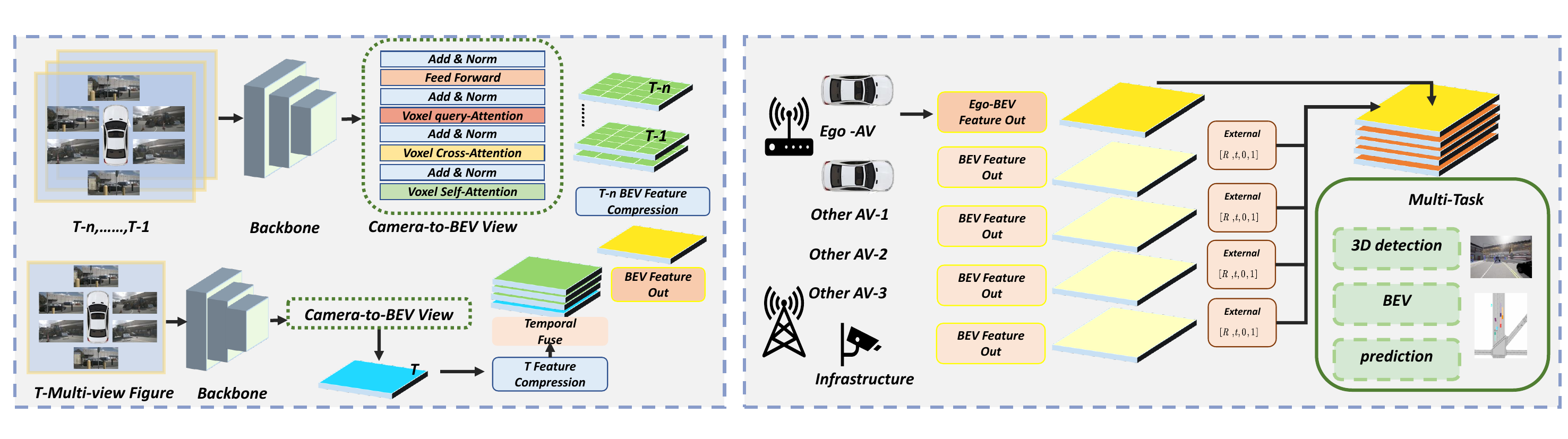}
  \caption{The V2X-perception architecture.}
  \label{fig:fig2}
\end{figure}

%	\subsection{Mission}
%	\label{Mission}
 
%我们对提出的任务进行准确的定义，以支持对下属模块的深入理解

%To better understand our work, we provide precise definitions for the proposed tasks. The primary tasks include the perception module and the interaction module with the LLM. For the perception module, the task involves generating 3D object detection boxes, BEV perception, and trajectory prediction through inputs from multiple devices and multiple perspectives. Regarding the natural language reasoning tasks, our main focus is to utilize the input and output information from the perception tasks as queried by the query module. We then construct a learnable and updatable LLM corpus for prompts. This facilitates proactive prompts through drivers' active inquiries and passive prompts from the model, offering drivers passive cues such as advance notices and warnings for traffic safety events. It also involves responding to drivers' active inquiries and addressing tasks initiated by traffic authorities, such as road traffic safety risk assessments and determining responsibilities in traffic accidents. This comprehensive approach aims to address various tasks related to traffic and vehicle safety in the context of autonomous driving and human-machine interaction, as shown in Figure \ref{fig:fig5}

	\subsection{V2X-perception}
	\label{V2X-perception}

We position the V2X-perception framework as a central component of AccidentGPT's perceptual work. This framework seamlessly integrates and extracts features from the panoramic multi-images captured by multiple vehicles and road-testing devices. Through a well-constructed deep learning framework, it efficiently produces accurate 3D detection boxes for vehicles, perception results in the BEV perspective, and trajectory prediction results. These outputs serve as crucial inputs for GPT-reasoning, making the framework a vital element of AccidentGPT's perceptual processes. The V2X-perception framework plays a pivotal role in enhancing the overall roadside traffic environment perception and map reconstruction. This will contribute to advancing collaborative safety driving for autonomous vehicles, long-range and blind-spot safety alerts for human-driven vehicles, human-machine interaction, and real-time assessments of traffic environments and accidents.

\begin{figure}[b]
  \centering

\includegraphics[width=1\linewidth]{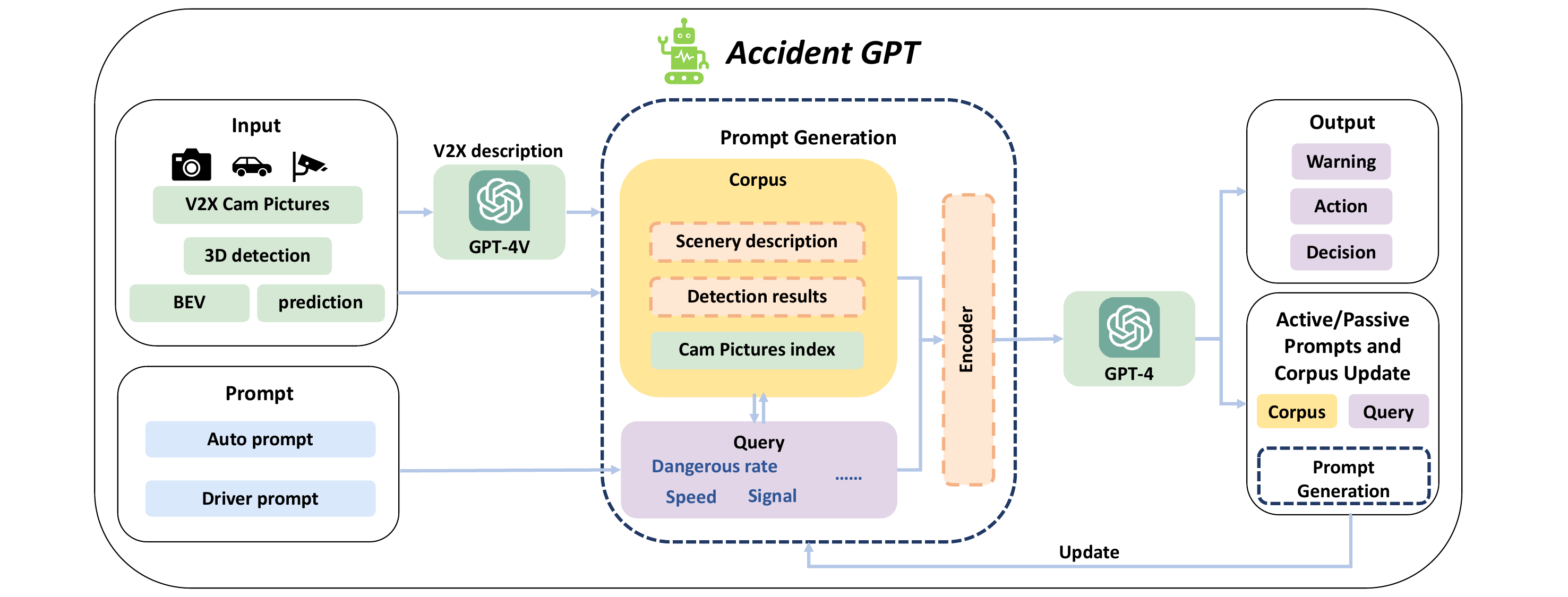}
  \caption{The GPT-reasoning module.}
  \label{fig:fig3}
\end{figure}

Specifically, in the realm of collaborative perception, drawing inspiration from BEVerse and DeepAccident, we have developed the global perception Module under the V2X architecture, referred to as V2X-perception as shown in Figure \ref{fig:fig2}. For the input data from images, we applied preprocessing to each image based on the image-backbone to extract diverse information and features. Subsequently, the information captured by the surround-view cameras in each vehicle or device is input into the cross transformer network. Additionally, we established a Camera-to-BEV based on camera parameters, facilitating the transformation of information from various vehicles or road testing devices into a BEV perspective.

To enhance the comprehensiveness of global perception and recognition accuracy, we align the perception information extracted at different time sequences for each vehicle and road testing device under the transformation of Ego-motion-T. This alignment results in the construction of a BEV feature out that represents the complete state.

To optimize the output types, efficiency, and quality of tasks, we devised multiple detection heads to achieve various task outputs. These include the conventional 3D object detection, BEV perspective map construction, and multi-vehicle trajectory prediction. These tasks uniformly employ BEV feature out for sampling and detection extraction, thereby enhancing the accuracy and completeness of multi-task perception work.
% 把任务的图片放到这里

	\subsection{GPT-reasoning}
	\label{GPT-reasoning}
 
The GPT-reasoning framework serves as the core element of AccidentGPT's functionality. Through intricate fusion and extraction of perceptual information along with various collected data, we introduce multiple modules within the semantic mega-model of GPT-4V. These modules efficiently provide advanced proactive cues for human-machine collaboration and passive information retrieval. The goal is to enhance the semantic interaction framework for human-machine collaboration, addressing both proactive and reactive tasks in traffic safety warnings and alerts. This, in turn, aims to elevate the bidirectional safety mission between humans and computers.

Specifically, in the V2X-perception module, the constructed model achieves comprehensive collaborative perception for multiple vehicles and road testing, producing outputs such as 3D object detection and trajectory prediction. This module provides a priori knowledge for GPT-reasoning, facilitating more efficient and accurate perceptual information support.

In the reasoning module, we define multiple decision tasks, establishing the multi-level priority sampling prompt module under corpus fusion, passive and active task prompt modules, and output modules for various accident-related tasks. This approach, supported by the standard GPT-4V framework, accurately perceives and describes decision tasks, offering advanced active and passive prompts to drivers or providing real-time road traffic environment assessments and accident causation analysis reports for traffic and planning departments. This helps reduce human-machine disparities, thereby enhancing human-vehicle interaction efficiency and ultimately reducing the occurrence of traffic accidents.

Specifically, the computational and interactive processes of our reasoning are illustrated in Figure \ref{fig:fig3}, encompassing input, query, corpus sampling, prompt generation, LLM computation, active/passive prompts, corpus updates, and output modules. We will now elaborate on the design of these modules, emphasizing their positive impact on enhancing human-machine interaction efficiency and providing proactive and reactive safety prompts, as well as analysis reports, under perceptual fusion.

\textbf{Input Module:} The input serves as the header of the entire module, covering all relevant information inputs. Within the specific module inputs, there are primarily two parts: the input of perceptual information and the input of corpus information. For the perceptual information part, it requires inputs that encompass results from perception, including target detection, BEV, and trajectory prediction information. Additionally, it includes labeled raw image information. Regarding the input of corpus information, it mainly consists of language suggestions provided by the driver, the textual content of previous conversations, and suggested queries for the constructed corpus. All the aforementioned input information is not directly analyzed and understood by the LLM; instead, it is initially input into an information repository for further extraction and retrieval.

\textbf{Query Module:} This segment is a pivotal element in the system, offering proactive remote safety warnings and blind-spot alerts for human-driven vehicles. It can also provide comprehensive intelligent traffic safety analysis and reports for traffic management and planning departments based on prompts and suggestions. Specially, it enhances safety by providing improved driving recommendations and behavioral guidelines through human-machine dialogue and interaction. Drivers can set threshold levels for proactive warnings through dialogue, activating the proactive warning feature to ensure safe driving. When proactive prompts are enabled, there is continuous sampling and processing of image data from the vehicle and its surroundings. This image data serves as pre-input into the LLM during the query process. The results of image analysis are stored in a temporary corpus, enabling subsequent queries and retrievals. Once the triggered information reaches a specific intensity set by the threshold, it activates proactive warnings and prompts to ensure the driver's safe operations. In the absence of proactive prompts, the query module remains inactive and instead passively collects various information based on the driver's commands and suggestions. This includes both image and textual data, serving as pre-input into the LLM. The captured data is stored in a temporary corpus to support further processing, providing results and suggestions to the user side.

\textbf{Corpus sampling module:} After the processing by the aforementioned input and query modules, various types of information are efficiently transformed into the desired textual descriptions. This transformation facilitates the precise matching of descriptive content with the predefined and constructed corpus. In this section, a predefined corpus module is established to organize diverse and multimodal information into the "repository" of the corpus. Within this repository, various data is packaged and organized into standardized "boxes," accurately converting descriptive data into precise descriptive text. This further enables subsequent modules to select the optimal input information for LLM, supporting LLM in better understanding the current traffic situation and accelerating its reasoning speed.

\textbf{Prompt Generation Module:} This section, through the integration of preliminary work, establishes methods and suggestions for generating prompts that assist the large model in better understanding the input content and the objectives of the current task, thereby enhancing the quality of inference results. Specifically, the prompt generation module comprises three key components: the original information backup query retrieval module, the temporal corpus construction module for the current moment, and the corpus sampling module, which includes compressed and refined information from the corpus sampling module. In the original information backup query retrieval module, all raw image and text-type information from the current and historical data are provided. This supplementation aims to fulfill the specific needs of the LLM's inference and prompt tasks for the current task. The information encompasses historical images, trajectories, vectorized map data, and their corresponding specific labels. Additionally, the temporal corpus construction module dynamically builds a temporary corpus for short-term or user-defined interaction expectations at the current moment. This acts as a dynamic "cache" of information, facilitating rapid interaction, extraction, and analysis of environmental and driver prompt information for subsequent inference steps. The corpus sampling module serves as the primary information source for LLM computation. It includes standardized "boxes" representing various sources and modalities of information. The corpus sampling module integrates and filters information from these "boxes," extracting suitable prompt frameworks. The output from this module is then incorporated into the original information backup query retrieval module and the temporal corpus construction module, ultimately providing integrated and constructed input information for posing questions to the large model.

\textbf{LLM Computation Module:}  In this phase, all prepared prompts are systematically input into the LLM to undergo intricate inference processes, expecting the LLM to make accurate decisions. Additionally, recognizing the inherent complexity and variability of driving scenarios and traffic safety environments, which may lead to unconventional biased decisions in semantic understanding, we adopt the chain-of-thought method to ensure the appropriate level of decision-making within the same input. This comprehensive approach integrates collaborative perception and reasoning to achieve improved human-vehicle interaction and contribute to advanced decision support systems for driving safety in intelligent connected vehicle environments. It also provides comprehensive recommendations for traffic management departments to monitor real-time road traffic safety risks.

\begin{figure}[h]
  \centering

\includegraphics[width=1\linewidth]{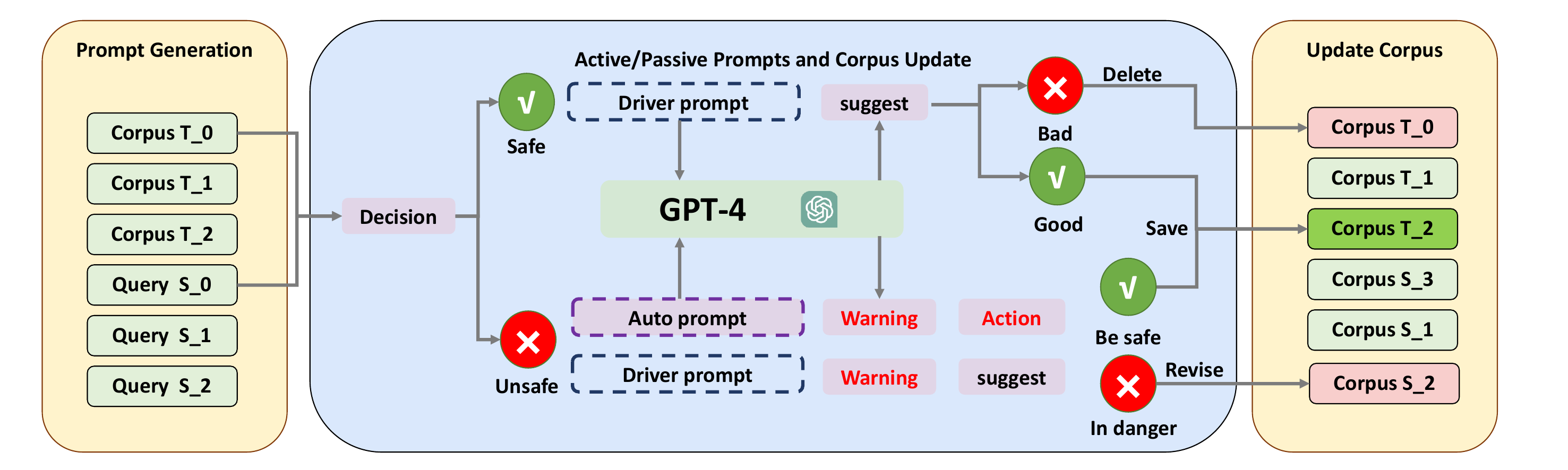}
  \caption{Active/Passive prompts and corpus update module.}
  \label{fig:fig4}
\end{figure}

\textbf{Active/Passive Prompts and Corpus Update Module:} In the process of providing proactive remote safety warnings and blind-spot alerts, as well as improved safety driving recommendations and behavioral guidelines through human-machine dialogue and interaction, we aim to implement two distinct sets of prompting rules for drivers, adjusting the prompt frequency to ensure safety margins. Specifically, within the active/passive prompts and corpus update module, we utilize the LLM to execute both proactive and reactive prompting tasks with the assistance of the proposed prompt module. The LLM evaluates whether the ongoing behavior aligns with the anticipated safety risk prompt range for the driver, enabling proactive prompts. The information prompted is then utilized as prior knowledge for subsequent prompts, accumulating valuable safety margin experience and enriching the corpus module. In situations where the system identifies significant safety risks, regardless of the driver's prompt requests, risk prompts need to be issued, potentially involving adaptive driving strategies. For corpus updates, we introduce an iterative update method that continuously learns from past accident cases and decisions, gradually enhancing the module's robustness and inference accuracy, elevating the inference capabilities to a higher dimension, as depicted in Figure \ref{fig:fig4}. Specifically, during a driving task, we record the prompts used based on the input from the driving scenario and the corresponding decisions generated by the LLM for each decision frame. Once the driving session concludes without collisions or hazardous events, signifying a successful completion, AccidentGPT samples key decision frames from the sequence. These frames become integral for assessing risk experience, thereby enriching the corpus module. Conversely, if the current scenario leads to a dangerous situation, such as a collision due to misjudgments and decisions, indicating a system decision bias, it is semantically recorded to improve future decision outcomes. This iterative process gradually achieves the objective of enabling the AccidentGPT system to learn from its mistakes.

\begin{figure}[h]
  \centering

\includegraphics[width=0.7\linewidth]{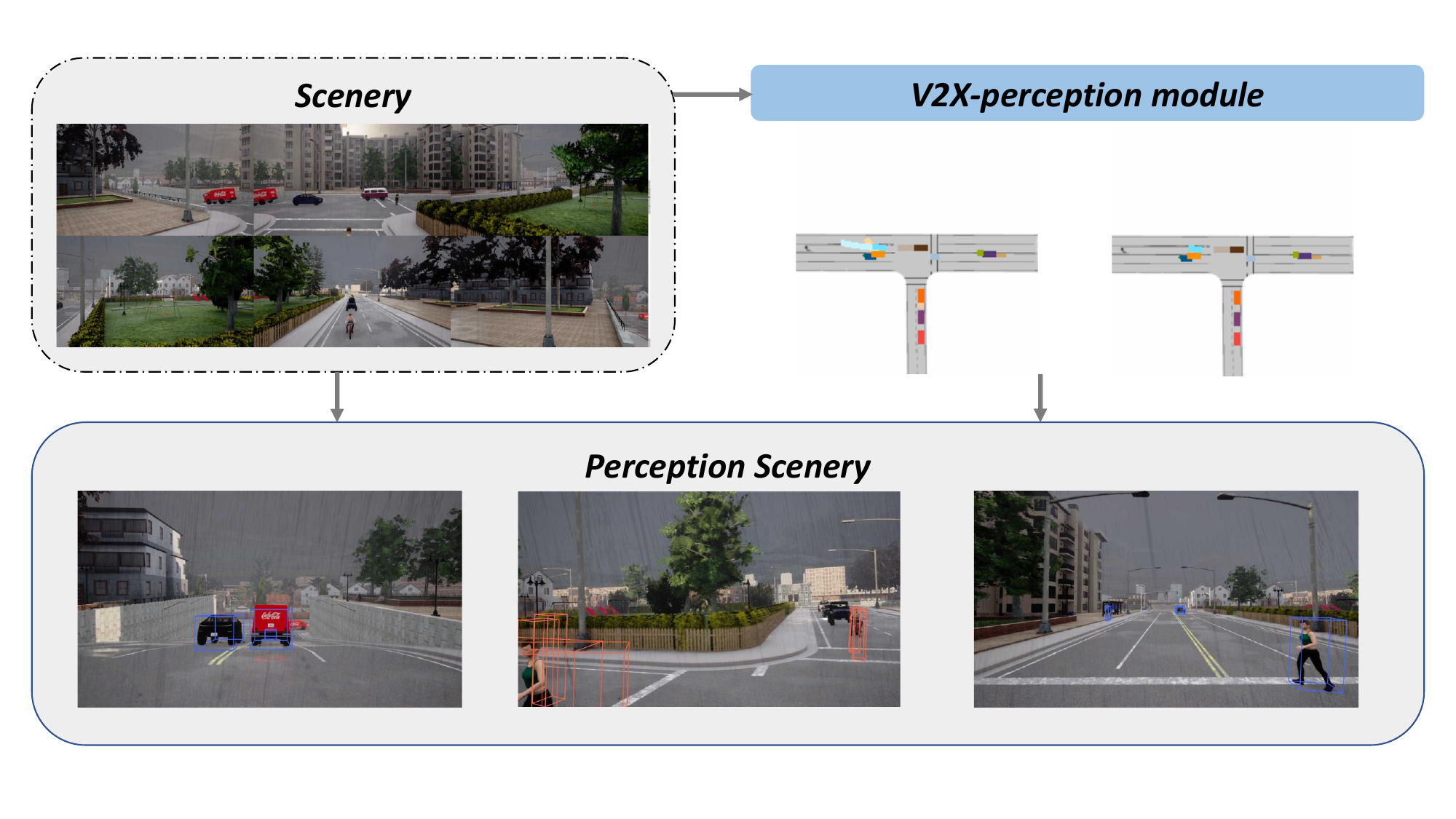}
  \caption{Automatic driving perception case study about AccidentGPT.}
  \label{fig:fig21}
\end{figure}

\begin{figure}[h]
  \centering

\includegraphics[width=0.7\linewidth]{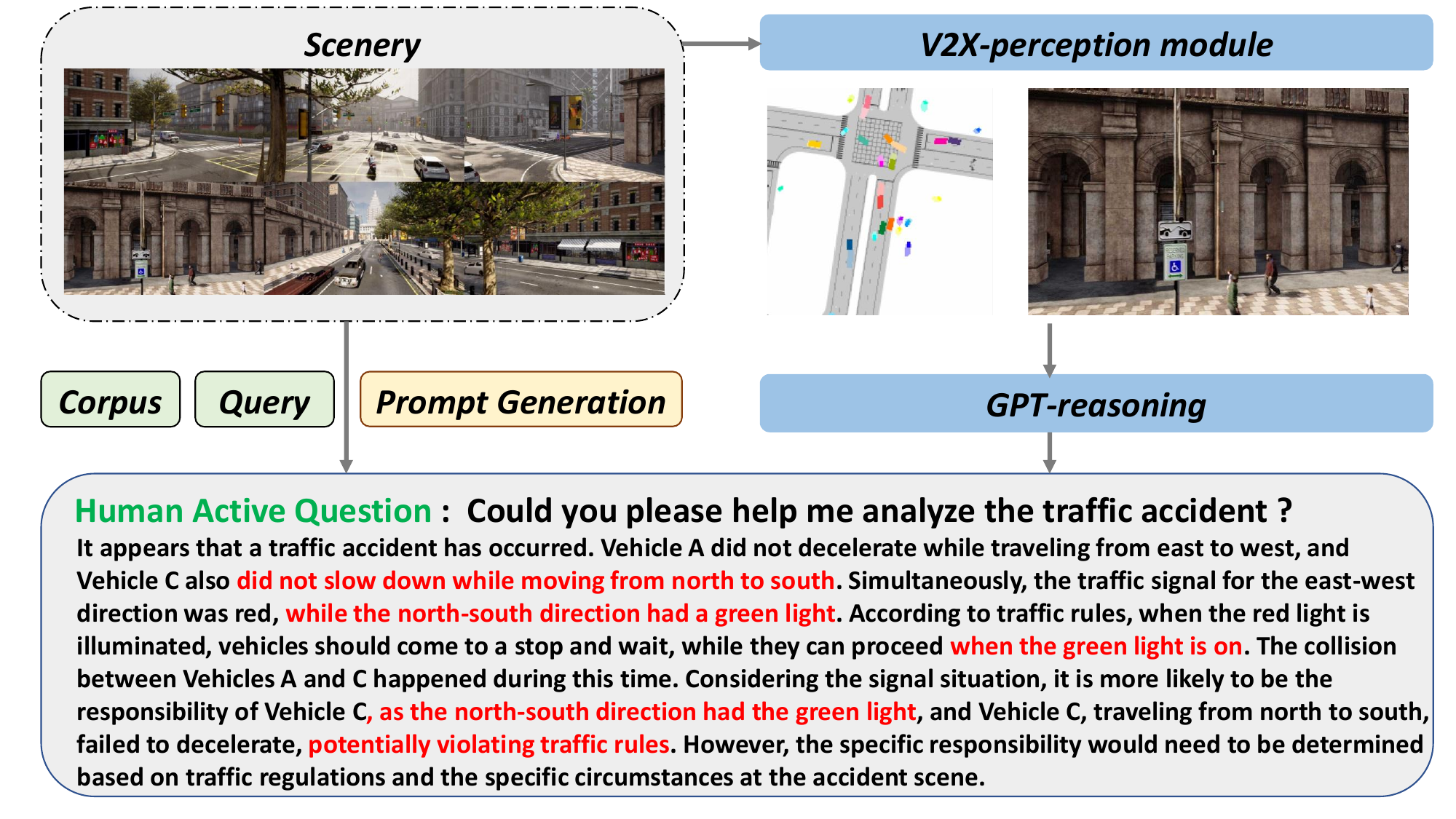}
  \caption{Active prompt case study about AccidentGPT.}
  \label{fig:fig19}
\end{figure}

\begin{figure}[h]
  \centering

\includegraphics[width=0.7\linewidth]{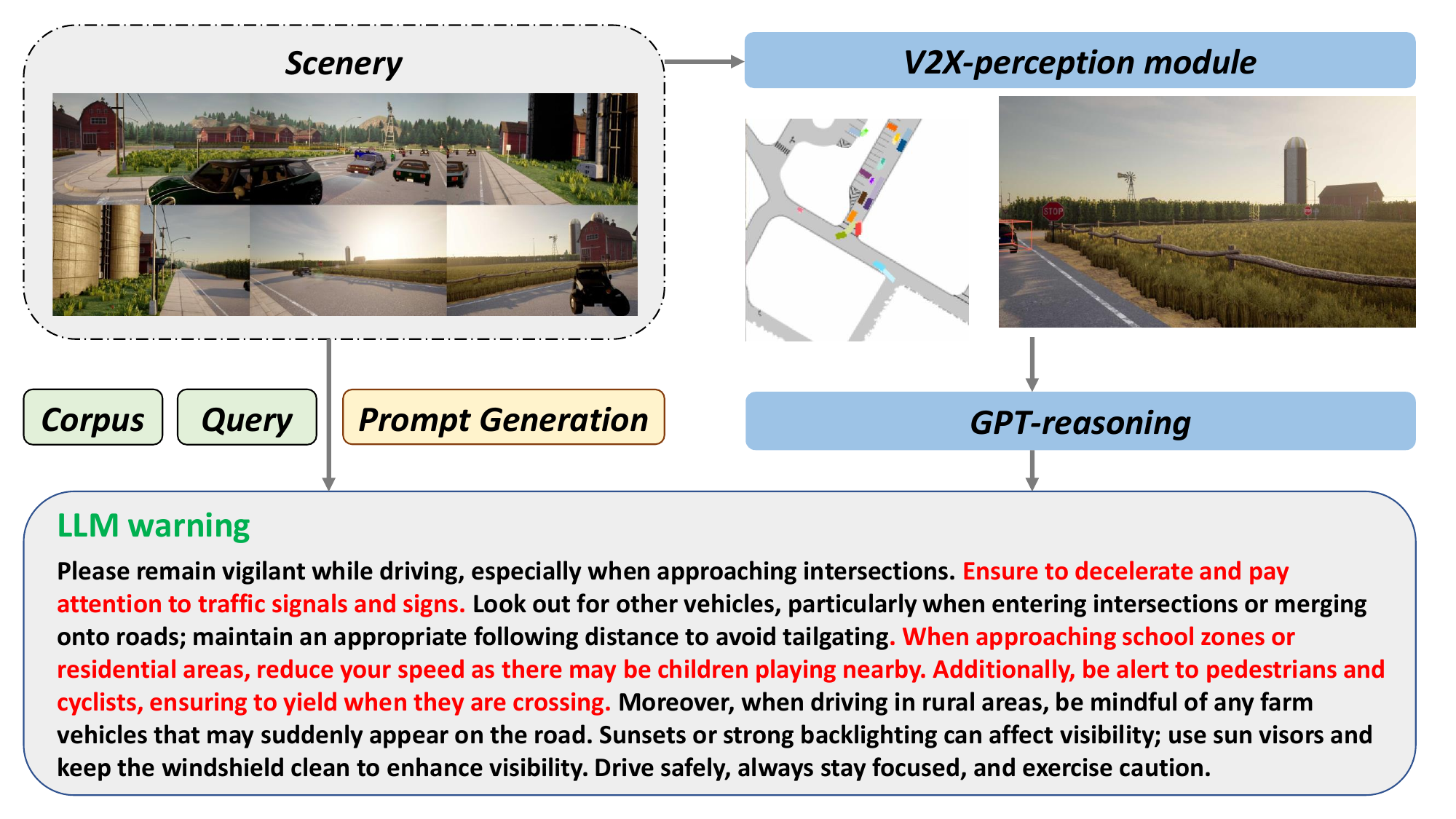}
  \caption{Passive prompt case study about AccidentGPT.}
  \label{fig:fig20}
\end{figure}

\begin{figure}[h]
  \centering

\includegraphics[width=0.7\linewidth]{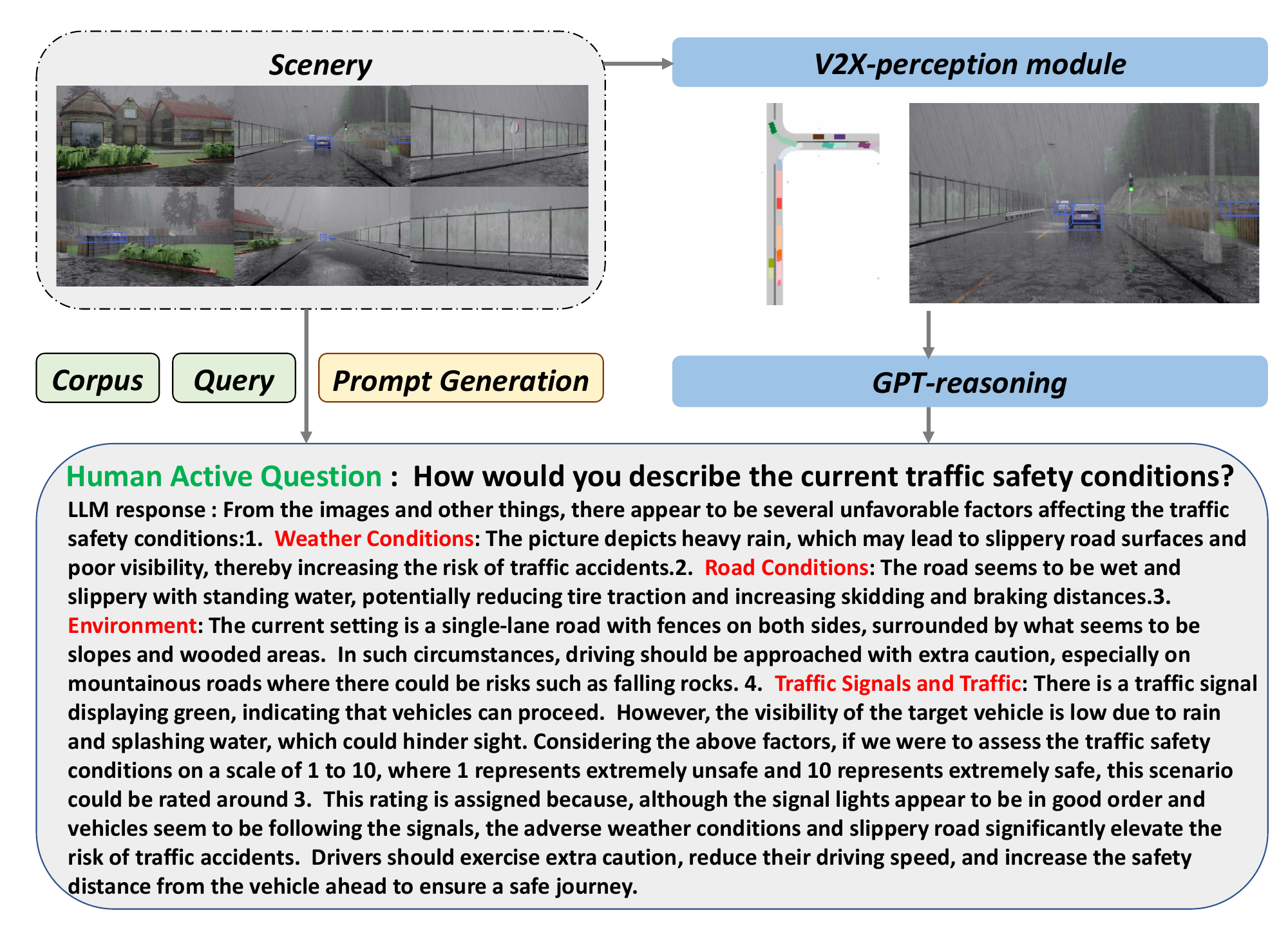}
  \caption{Traffic environment analysis case study about AccidentGPT.}
  \label{fig:fig22}
\end{figure}

\textbf{Output Module:}Ultimately, the process involves extracting semantic content from the LLM representation and executing decisions and actions. The interconnection of each previously mentioned module culminates in the final decision, decoded from the feedback decoder in the LLM, thereby establishing a GPT-driven reasoning system. This comprehensive approach, incorporating both active and passive prompting, facilitates continuous learning through corpus updates and relies on LLM-driven decision-making.

This significantly contributes to the development of an advanced AccidentGPT system with adaptive learning capabilities, enabling effective interactive advice, tips, and traffic safety alerts to guide safety decisions in human-machine hybrid driving scenarios until fully autonomous driving is achieved. It also benefits traffic police and management agencies by enabling real-time human-machine interaction through the collaborative perception of global vehicles and road devices. This leads to intelligent analysis of the overall traffic safety environment and a comprehensive examination of the causes and responsibilities of accidents based on historical data from vehicle collisions.

\section{Experimental results}
\label{Experimental results}

	\subsection{Case study}
	\label{Case study}

To gain a deeper understanding of our work, this section provides three case study illustrating the primary functionalities of AccidentGPT.

In Figure \ref{fig:fig21}, for autonomous vehicles, the model performs comprehensive environmental perception and understanding based on individual vehicle intelligence, vehicle-to-vehicle coordination, and vehicle-to-road coordination. It achieves precise scene perception, target detection, and trajectory prediction, enabling effective collision avoidance.

\begin{figure}[h]
  \centering

\includegraphics[width=0.7\linewidth]{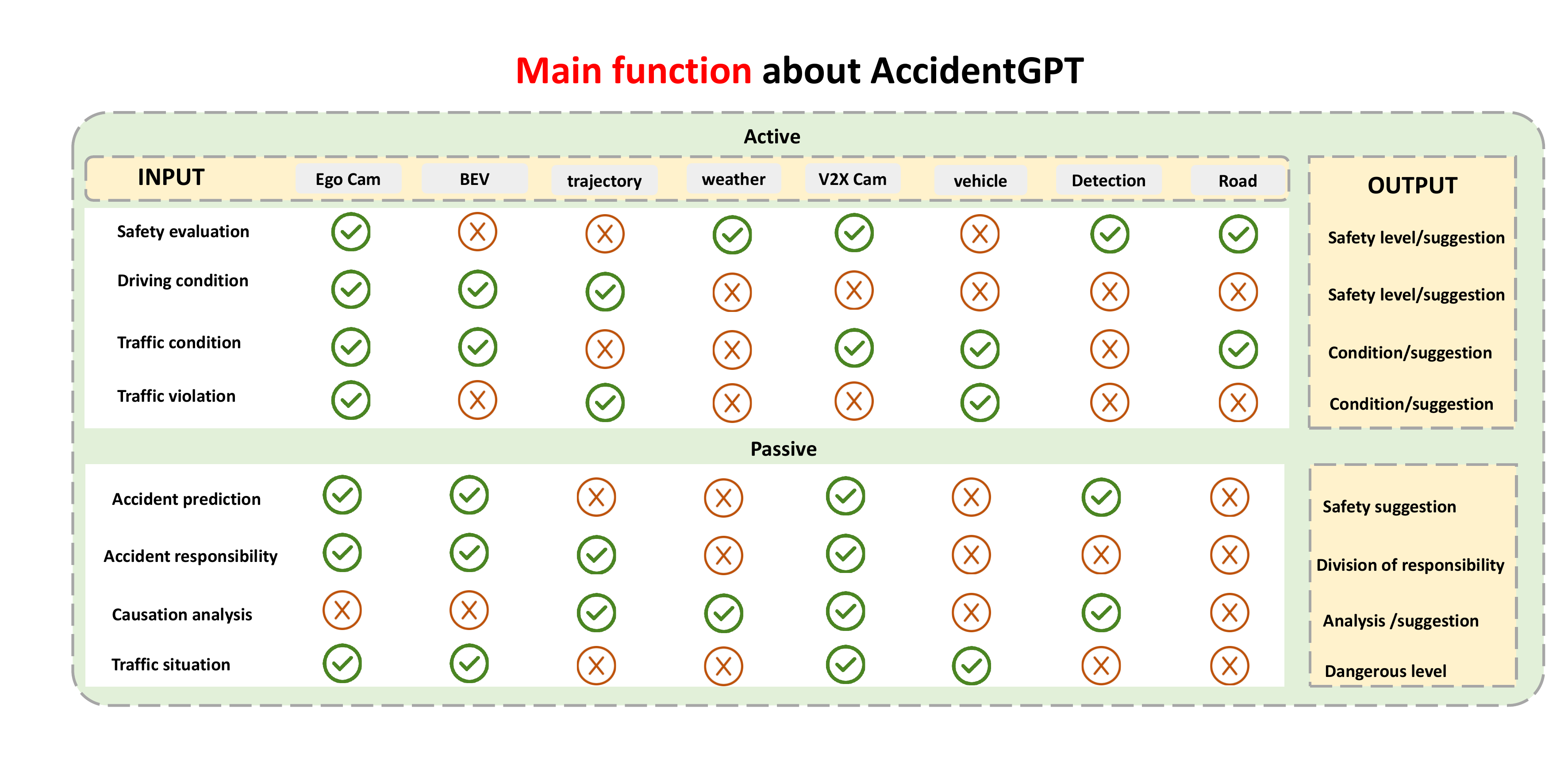}
  \caption{Main function about AccidentGPT.}
  \label{fig:fig5}
\end{figure}

For human-driven vehicles, as shown in Figure \ref{fig:fig19}, the driver actively inquires about the traffic situation. The AccidentGPT model utilizes the V2X perception module to interpret images, combines it with GPT reasoning network, and offers the driver judgments and predictions regarding vehicles, roads, and specific landmarks. The outcome of this interaction is providing warnings and suggestions to the driver. In Figure \ref{fig:fig20}, guided by the safety alert threshold set by the driver in the previous dialogue, when the vehicle approaches a complex intersection or area, the large model takes a proactive stance, providing safety driving recommendations. It suggests the driver to exercise caution at signal-less intersections and emphasizes the visibility challenges posed by backlit conditions during dusk. This proactive approach aims to issue timely warnings to the driver based on the current environmental conditions. Therefore, it can provide proactive long-range safety warnings and blind-spot alerts to the driver and offer better safety driving recommendations and behavioral guidelines through human-machine dialogue and interaction.

\begin{table}[h]
\centering
\caption{Comparison of detection and motion results.}
\renewcommand{\arraystretch}{1.5} % 调整行间距
\begin{tabular}{l|llll}
\hline
\multicolumn{1}{c|}{}                                    & \multicolumn{3}{c}{Index}                                                                      \\ \cline{2-4} 
\multicolumn{1}{c|}{}                                    & \multicolumn{1}{c|}{}                            & \multicolumn{2}{c}{Motion}                  \\ \cline{3-4} 
\multicolumn{1}{c|}{\multirow{-3}{*}{Different   model}} & \multicolumn{1}{c|}{\multirow{-2}{*}{Detection (\%) }} & mIOU (\%)                        & VPQ (\%)           \\ \hline
Average Fusion  \cite{wang2023deepaccident}                                         & 36.2                                             & { 52.1} & 39.5          \\
DiscoNet \cite{mehr2019disconet}                                              & 38.5                                             & 54.2                        & 42            \\
V2X-ViT  \cite{xu2022v2x}                                                & 40.1                                             & 55.1                        & 43.2          \\
CoBEVT    \cite{xu2022cobevt}                                               & 40.8                                             & 56.2                        & 44            \\
V2X-perception (our)                                      & \textbf{41.07}                                   & \textbf{57.3}              & \textbf{45.2} \\ \hline
\end{tabular}
\label{tab:table004}
\end{table}

\begin{table}[h]
\centering
\caption{Comparison of different type results.}
\renewcommand{\arraystretch}{1.5} % 调整行间距
\begin{tabular}{l|llll}
\hline
\multicolumn{1}{c|}{\multirow{2}{*}{Test   model type}} & \multicolumn{4}{c}{Metric}                    \\ \cline{2-5} 
\multicolumn{1}{c|}{}                                   & mATE (m)        & mASE (1-IoU)    & mAOE (rad)    & mAVE (m/s)     \\ \hline
Car                                                     & 0.994540292 & 0.964026 & 1.022296 & 1.267359 \\
Truck                                                   & 0.995231552 & 0.957066 & 1.01609  & 1.205831 \\
Van                                                     & 0.997434757 & 0.986803 & 1.012704 & 1.103993 \\
Pedestrian                                              & 1.001661385 & 0.999283 & 1.025322 & 1.026851 \\ \hline
\end{tabular}
\label{tab:table005}
\end{table}

\begin{figure}[h]
  \centering

\includegraphics[width=0.7\linewidth]{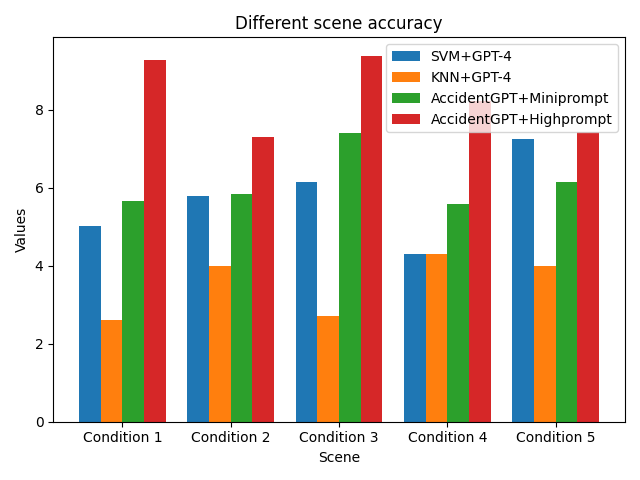}
  \caption{Comparison of partition Models in different scenarios' outcomes.}
  \label{fig:fig7}
\end{figure}

\begin{figure}[h]
  \centering

\includegraphics[width=0.7\linewidth]{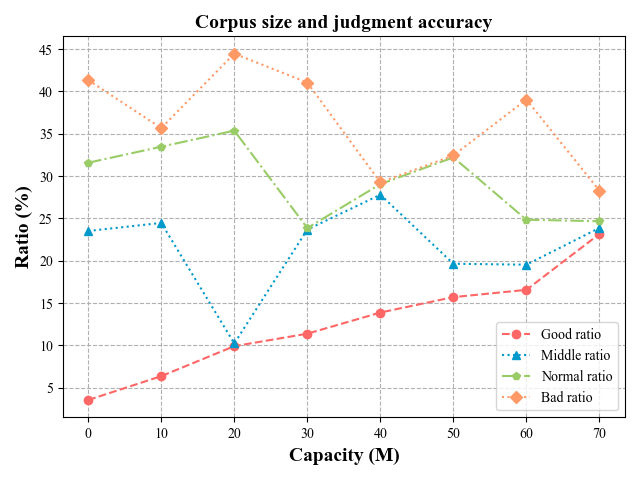}
  \caption{Comparison of AccidentGPT models in different corpus size.}
  \label{fig:fig6}
\end{figure}

\begin{table}[h]
\centering
\caption{Comparison of different partition model results}
\renewcommand{\arraystretch}{1.5} % 调整行间距
\begin{tabular}{c|lllll}
\hline
                                       & \multicolumn{1}{c}{}                       & \multicolumn{4}{c}{Index}                                                                                                            \\ \cline{3-6} 
\multirow{-2}{*}{Partition model}      & \multicolumn{1}{c}{\multirow{-2}{*}{Type}} & {Accuracy (\%) } & {Precision (\%) } & { Recall (\%) } & {F1-score (\%) } \\ \hline
                                       & Scene                                      & 0.53                            & 0.42                             & 0.34                          & 0.34                            \\
                                       & Safety                                     & 0.74                            & 0.78                             & 0.78                          & 0.74                            \\
\multirow{-3}{*}{SVM+GPT-4 \cite{joachims1998making}}        & Accident                                   & 0.74                            & 0.51                             & 0.5                           & 0.49                            \\ \hline
                                       & Scene                                      & 0.47                            & 0.29                             & 0.3                           & 0.26                            \\
                                       & Safety                                     & 0.16                            & 0.15                             & 0.23                          & 0.16                            \\
\multirow{-3}{*}{KNN+GPT-4 \cite{guo2003knn} }        & Accident                                   & 0.47                            & 0.3                              & 0.3                           & 0.27                            \\ \hline
                                       & Scene                                      & \textbf{0.89}                   & \textbf{0.83}                    & \textbf{0.83}                 & \textbf{0.84}                   \\
                                       & Safety                                     & \textbf{0.89}                   & \textbf{0.92}                    & \textbf{0.89}                 & \textbf{0.89}                   \\
\multirow{-3}{*}{AccidentGPT (GPT-reasoning+GPT-4)} & Accident                                   & \textbf{0.89}                   & \textbf{0.92}                    & \textbf{0.83}                 & \textbf{0.85}                   \\ \hline
\multicolumn{1}{l}{}                   &                                            &                                 &                                  &                               &                                 \\
\multicolumn{1}{l}{}                   &                                            &                                 &                                  &                               &                                
\end{tabular}
\label{tab:table001}

\end{table}

\begin{table}[h]
\centering
\caption{Comparison of AccidentGPT models in different prompt.}
\renewcommand{\arraystretch}{1.4} % 调整行间距
\begin{tabular}{c|llllll}
\hline
                                   & \multicolumn{1}{c|}{}                          & \multicolumn{5}{c}{Index}                                                                                                                                                \\ \cline{3-7} 
\multirow{-2}{*}{Different prompt} & \multicolumn{1}{c|}{\multirow{-2}{*}{Misson}}  & \multicolumn{1}{c}{Number}                    & {Good (\%)} & {Middle (\%) } & {Normal (\%) } & {Bad (\%) } \\ \hline
                                   & {   Safety evaluation}       & \multicolumn{1}{l|}{{   1}} & 2.61                        & 26.18                         & 32.08                         & 39.12                      \\
                                   & {   Driving condition}       & \multicolumn{1}{l|}{{   2}} & 5.62                        & 13.08                         & 26.24                         & 55.06                      \\
                                   & {   Traffic condition}       & \multicolumn{1}{l|}{{   3}} & 3.81                        & 7.76                          & 35.79                         & 52.64                      \\
                                   & {   Traffic violation}       & \multicolumn{1}{l|}{{   4}} & 4.12                        & 28.19                         & 32.65                         & 35.04                      \\
                                   & {   Accident prediction}     & \multicolumn{1}{l|}{{   5}} & 3.33                        & 21.59                         & 31.78                         & 43.3                       \\
                                   & {   Accident responsibility} & \multicolumn{1}{l|}{{   6}} & 3.99                        & 10.77                         & 41.2                          & 44.04                      \\
                                   & {   Causation analysis}      & \multicolumn{1}{l|}{{   7}} & 5.2                         & 21.58                         & 34.37                         & 38.85                      \\
\multirow{-8}{*}{Mini+prompt}      & {   Traffic situation}       & \multicolumn{1}{l|}{{   8}} & 4.52                        & 5.2                           & 43.39                         & 46.89                      \\ \hline
                                   & {   Safety evaluation}       & \multicolumn{1}{l|}{{   1}} & 7.43                        & 16.21                         & 32.6                          & 43.76                      \\
                                   & {   Driving condition}       & \multicolumn{1}{l|}{{   2}} & 7.85                        & 23.57                         & 30.32                         & 38.26                      \\
                                   & {   Traffic condition}       & \multicolumn{1}{l|}{{   3}} & 7.76                        & 10.97                         & 39.06                         & 42.2                       \\
                                   & {   Traffic violation}       & \multicolumn{1}{l|}{{   4}} & 10.24                       & 19.46                         & 25.79                         & 44.51                      \\
                                   & {   Accident prediction}     & \multicolumn{1}{l|}{{   5}} & 10.87                       & 27.05                         & 27.18                         & 34.89                      \\
                                   & {   Accident responsibility} & \multicolumn{1}{l|}{{   6}} & 8.32                        & 25.34                         & 25.42                         & 40.92                      \\
                                   & {   Causation analysis}      & \multicolumn{1}{l|}{{   7}} & 8.46                        & 19.35                         & 36.05                         & 36.14                      \\
\multirow{-8}{*}{Middle+prompt}    & {   Traffic situation}       & \multicolumn{1}{l|}{{   8}} & 7.14                        & 22.15                         & 33.87                         & 36.84                      \\ \hline
                                   & {   Safety evaluation}       & \multicolumn{1}{l|}{{   1}} & 10.83                       & 16.34                         & 31.87                         & 40.96                      \\
                                   & {   Driving condition}       & \multicolumn{1}{l|}{{   2}} & 13.03                       & 15.46                         & 31.74                         & 39.76                      \\
                                   & {   Traffic condition}       & \multicolumn{1}{l|}{{   3}} & 16.41                       & 19.54                         & 31.72                         & 32.33                      \\
                                   & {   Traffic violation}       & \multicolumn{1}{l|}{{   4}} & 15.72                       & 20.92                         & 27.31                         & 36.05                      \\
                                   & {   Accident prediction}     & \multicolumn{1}{l|}{{   5}} & 9.42                        & 29.17                         & 29.83                         & 31.58                      \\
                                   & {   Accident responsibility} & \multicolumn{1}{l|}{{   6}} & 17.56                       & 25.08                         & 25.97                         & 31.39                      \\
                                   & {   Causation analysis}      & \multicolumn{1}{l|}{{   7}} & 16.62                       & 19.03                         & 32.08                         & 32.26                      \\
\multirow{-8}{*}{High+prompt}      & {   Traffic situation}       & \multicolumn{1}{l|}{{   8}} & 13.12                       & 17.8                          & 21.18                         & 47.9                       \\ \hline
\end{tabular}
\label{tab:table003}
\end{table}

In Figure \ref{fig:fig22}, for traffic police and management agencies, intelligent and real-time traffic safety analysis and comprehensive evaluation reports regarding pedestrian, vehicles, roads, and the environment can be achieved through collaborative perception from multiple vehicles and road testing devices. This includes in-depth analysis of traffic safety, providing a holistic assessment and analysis of accident causes and responsibilities following vehicle collisions.

\begin{table}[p]
\centering
\caption{Comparison of AccidentGPT models in different prompt renewal rate.}
\renewcommand{\arraystretch}{1.4} % 调整行间距
\begin{tabular}{c|llllll}
\hline
                                        & \multicolumn{1}{c|}{}                            & \multicolumn{5}{c}{Index}                                                                                                                             \\ \cline{3-7} 
\multirow{-2}{*}{Renewal   rate}        & \multicolumn{1}{c|}{\multirow{-2}{*}{Misson}}    & \multicolumn{1}{c}{Number} & {Good (\%) } & { Middle (\%) } & {Normal (\%) } & {Bad (\%) } \\ \hline
                                        & { Safety evaluation}         & { 1}   & 2.87                        & 22.81                         & 31.7                          & 42.62                      \\
                                        & { Driving   condition}       & { 2}   & 4.82                        & 20.49                         & 35.98                         & 38.71                      \\
                                        & { Traffic   condition}       & {  3}   & 3.56                        & 23.52                         & 31.58                         & 41.34                      \\
                                        & { Traffic   violation}       & { 4}   & 6.46                        & 13.33                         & 22.01                         & 58.19                      \\
                                        & { Accident   prediction}     & { 5}   & 6.07                        & 11.15                         & 41.36                         & 41.42                      \\
                                        & { Accident   responsibility} & { 6}   & 4.76                        & 8.63                          & 32.04                         & 54.57                      \\
                                        & { Causation   analysis}      & { 7}   & 3.3                         & 3.97                          & 34.28                         & 58.44                      \\
\multirow{-8}{*}{10\% Renewal prompt}   & { Traffic   situation}       & { 8}   & 2.53                        & 12.63                         & 41.06                         & 43.79                      \\ \hline
                                        & { Safety evaluation}         & { 1}   & 9.05                        & 25.79                         & 26.08                         & 39.08                      \\
                                        & { Driving condition}         & { 2}   & 2.38                        & 14.89                         & 40.1                          & 42.63                      \\
                                        & {  Traffic condition}         & {  3}   & 4.34                        & 24.9                          & 31.7                          & 39.06                      \\
                                        & {  Traffic violation}         & {  4}   & 6.38                        & 24.47                         & 33.47                         & 35.67                      \\
                                        & {  Accident prediction}       & {  5}   & 9.73                        & 21.03                         & 22.6                          & 46.64                      \\
                                        & {  Accident responsibility}   & {  6}   & 6.77                        & 19.08                         & 19.93                         & 54.21                      \\
                                        & {  Causation analysis}        & {  7}   & 6.49                        & 27.49                         & 29.9                          & 36.11                      \\
\multirow{-8}{*}{30\% Renewal prompt}   & {  Traffic situation}         & {  8}   & 9.92                        & 10.24                         & 35.37                         & 44.46                      \\ \hline
                                        & {  Safety evaluation}         & {  1}   & 12.46                       & 16.07                         & 26.15                         & 45.31                      \\
                                        & {  Driving condition}         & {  2}   & 11.39                       & 23.69                         & 23.87                         & 41.05                      \\
                                        & {  Traffic condition}         & {  3}   & 14.1                        & 25.88                         & 27.31                         & 32.71                      \\
                                        & {  Traffic violation}         & {  4}   & 19.45                       & 21.17                         & 24.69                         & 34.69                      \\
                                        & {  Accident prediction}       & {  5}   & 15.87                       & 16.03                         & 33.25                         & 34.86                      \\
                                        & {  Accident responsibility}   & {  6}   & 16.47                       & 24.68                         & 29.09                         & 29.76                      \\
                                        & {  Causation analysis}        & {  7}   & 13.89                       & 27.8                          & 29.05                         & 29.26                      \\
\multirow{-8}{*}{50\%   Renewal prompt} & {  Traffic situation}         & {  8}   & 11.16                       & 26.63                         & 30.06                         & 32.15                      \\ \hline
                                        & {  Safety evaluation}         & {  1}   & 13.68                       & 19.77                         & 19.94                         & 46.61                      \\
                                        & {  Driving condition}         & {  2}   & 10.47                       & 25.15                         & 26.76                         & 37.61                      \\
                                        & {  Traffic condition}         & {  3}   & 16.31                       & 25.23                         & 28.7                          & 29.75                      \\
                                        & {  Traffic violation}         & {  4}   & 16.57                       & 19.54                         & 24.85                         & 39.04                      \\
                                        & {  Accident prediction}       & {  5}   & 11.16                       & 13.96                         & 20.28                         & 54.6                       \\
                                        & {  Accident responsibility}   & {  6}   & 18.35                       & 22.34                         & 24.96                         & 34.34                      \\
                                        & {  Causation analysis}        & {  7}   & 15.71                       & 19.65                         & 32.21                         & 32.44                      \\
\multirow{-8}{*}{70\% Renewal prompt}   & {  Traffic situation}         & {  8}   & 11.44                       & 15.11                         & 19.53                         & 53.93                      \\ \hline
                                        & {  Safety evaluation}         & {  1}   & 23.17                       & 23.88                         & 24.67                         & 28.28                      \\
                                        & {  Driving condition}         & {  2}   & 20.76                       & 22.5                          & 26.28                         & 30.47                      \\
                                        & {  Traffic condition}         & {  3}   & 20.26                       & 22.29                         & 28.51                         & 28.95                      \\
                                        & {  Traffic violation}         & {  4}   & 20.34                       & 22.14                         & 26.22                         & 31.31                      \\
                                        & {  Accident prediction}       & {  5}   & 20.34                       & 21.81                         & 26.53                         & 31.32                      \\
                                        & {  Accident responsibility}   & {  6}   & 15.95                       & 16.77                         & 22.82                         & 44.46                      \\
                                        & {  Causation analysis}        & {  7}   & 14.85                       & 27.03                         & 28.63                         & 29.49                      \\
\multirow{-8}{*}{90\% Renewal prompt}   & {  Traffic situation}         & {  8}   & 14.77                       & 22.44                         & 27.25                         & 35.54                      \\ \hline
\end{tabular}
\label{tab:table002}
\end{table}

	\subsection{Settings}
    \label{Settings}

In the experimental section, we will conduct tests and experiments on the proposed model to demonstrate its effectiveness and superiority. For perception and planning tasks, we will perform tests based on the DeepAccident dataset. This dataset includes perception results from six devices, comprising 4 cars' surround-view camera information and two road testing devices. The environment includes collisions with illegal activities such as running red lights and unexpected collisions during driving. Additionally, it contains various types of intersections and weather conditions for selection and evaluation. For testing based on LLM, we utilized OpenAI's GPT-4 for semantic tasks restructuring and integration, and GPT-4V for continuous frame image recognition and analysis. In our constructed AccidentGPT framework, we also established a corpus module for updating and learning to facilitate semantic and encoding transformations. In the experimental part, we will test and validate V2X-perception from aspects like image perception and prediction to demonstrate its superiority in 3D object detection, BEV perception, and trajectory prediction. For the GPT-reasoning module, we will evaluate its performance on eight different mainstream tasks, as shown in Figure \ref{fig:fig5}. We will conduct separate assessments and comparisons with benchmark methods and mainstream classification models. Ablation experiments will also be carried out to demonstrate the beneficial value and effectiveness of each proposed module.

	\subsection{Perception and planning}
    \label{perception and planning}

\textbf{Perception and Trajectory Prediction Performance.} As shown in Table \ref{tab:table004}, we conducted tests and validations by comparing our proposed V2X-perception model with other mainstream collaborative models, namely DiscoNet\cite{mehr2019disconet}, V2X-ViT \cite{xu2022v2x}, and CoBEVT \cite{xu2022cobevt}. It is evident that the V2X-perception model outperforms previous modules in both 3D target detection and trajectory prediction tasks, as judged by the three evaluation metrics. It consistently achieved the best performance, thereby providing more beneficial guidance for the subsequent GPT-reasoning module in terms of perception.

\textbf{Performance Based on Different Types of Vehicle Visibility.} During observation, the visibility of different types of vehicles or pedestrians is a crucial aspect of perception. Therefore, we simultaneously assessed the perceptual model's visibility differences for various vehicle or pedestrian types, as shown in Table \ref{tab:table005}. Specifically, we calculated the mean Average Precision (mAP) for each category, which measures the average accuracy; mean Average Surface Error (mASE), assessing the average error between the detected target surface and the actual surface; mean Average Orientation Error (mAOE), evaluating the average error between the detected target orientation and the actual orientation; and mean Average Velocity Error (mAVE), measuring the average error between the predicted and actual velocities. The results indicate that our model (V2X-perception) performs well in the perception of various vehicles and pedestrians, achieving a commendable level of effectiveness.

	\subsection{GPT-reasoning}
	\label{GPT-reasoning}

\textbf{Model performance in defining and classifying events.} To assess the performance of event definition, we compared our proposed end-to-end AccidentGPT method with the classification efficiency of SVM \cite{joachims1998making}, KNN \cite{guo2003knn} pre-processing combined with AccidentGPT. We conducted the analysis using four evaluation metrics:

\begin{itemize}
    \item  Accuracy:The proportion of correctly classified samples among all samples.

    \item  Precision:The proportion of true positive samples among those classified as a particular category.

    \item  Recall:The proportion of correctly classified samples among all samples belonging to a particular category.

    \item  F1 Score:The harmonic mean of precision and recall.
    
\end{itemize}
Regardless of the scenario definition, traffic risk delineation, or even accident responsibility assignment, the results obtained by conventional deep learning models were consistently lower than our proposed end-to-end AccidentGPT method, as shown in Table \ref{tab:table001}. We selected a specific scenario task and compared the SVM+AccidentGPT and KNN+AccidentGPT methods. These methods did not effectively judge and analyze images, leading to significant errors in subsequent semantic analysis. In contrast, the end-to-end AccidentGPT method more effectively captured accurate information, enabling precise judgment and decision-making, as illustrated in Figure \ref{fig:fig7}.

\textbf{Performance of Prompt Generation Intensity in Different Scenarios.} The intensity of prompt generation can significantly influence the accuracy of the model's output. We conducted experiments to analyze the impact of prompts with different intensities, categorizing the content volume of prompt generation as (Mini/Middle/High+Prompt), as illustrated in Table \ref{tab:table003}. It is evident that as the prompt generation intensity gradually increases, the performance of prompt results also improves across different scenarios.

\textbf{Performance of Model's Prompt Library Update Frequency Ratio on Model Inference.} The effectiveness of the model is, to a certain extent, attributed to the proposed prompt library update module. When the library is gradually updated based on feedback, it effectively provides deeper and more accurate guiding prompts for further semantic information. As indicated in Table \ref{tab:table002}, our analysis of prompt library update frequency reveals that as the library undergoes in-depth updates along with prompts, the accuracy of prompts in different corresponding scenarios will progressively improve. This improvement is particularly remarkable in the "good" category. In Figure \ref{fig:fig6}, the variation in prompt update capacity further underscores this point.

\section{Conclusion}
In order to establish a unified framework for a comprehensive understanding and application of traffic safety, aiming to enhance road traffic safety during the transition from manual to fully autonomous driving, this paper introduces an end-to-end accident analysis and prevention from V2X environmental perception with multi-modal large model AccidentGPT. This framework achieves three-dimensional object detection, BEV perception, and vehicle trajectory prediction across the entire road environment by constructing a comprehensive perception model under the V2X architecture. The integration of a natural language reasoning module with perception methods is designed to blend cognitive reasoning abilities with perceptual approaches. This integration is aimed at addressing various challenges in accident analysis and prevention within the field of traffic safety. Extensive experimental validation demonstrates that each proposed module exhibits significant potential and robust generalization capabilities in perception and safety tasks. Furthermore, AccidentGPT seamlessly integrates with various typical autonomous driving models and datasets, emphasizing its deployment potential in human-machine hybrid driving scenarios and traffic safety optimization and governance systems. To the best of our knowledge, we are the first to research and develop a large-scale model that combines perception with human cognition, integrating comprehensive scene perception and human understanding into traffic safety research. The demonstrated capabilities highlight its long-term application potential from current manual driving scenarios to the complete transition to autonomous driving systems, aiming to improve efficiency and ultimately reduce the occurrence of traffic accidents.

\section*{Data Availability}

Data page: \url{https://deepaccident.github.io}

%Bibliography
\bibliographystyle{unsrt}  
\bibliography{references}  

\newpage
\section*{Appendix}

\begin{figure}[h]
  \centering

\includegraphics[width=0.7\linewidth]{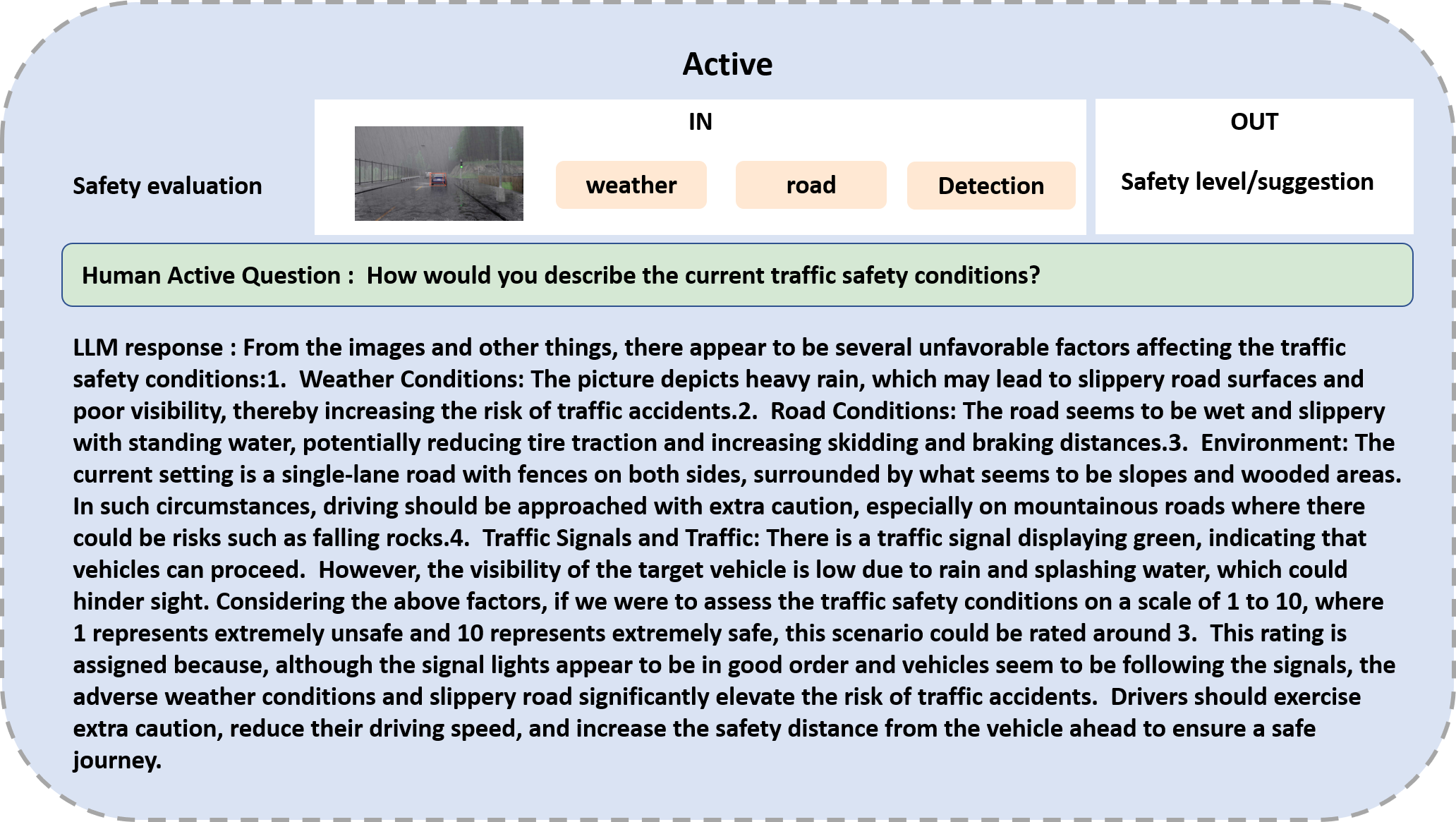}
  \caption{Sample of safety evaluation in AccidentGPT.}
  \label{fig:app01}
\end{figure}

\begin{figure}[h]
  \centering

\includegraphics[width=0.7\linewidth]{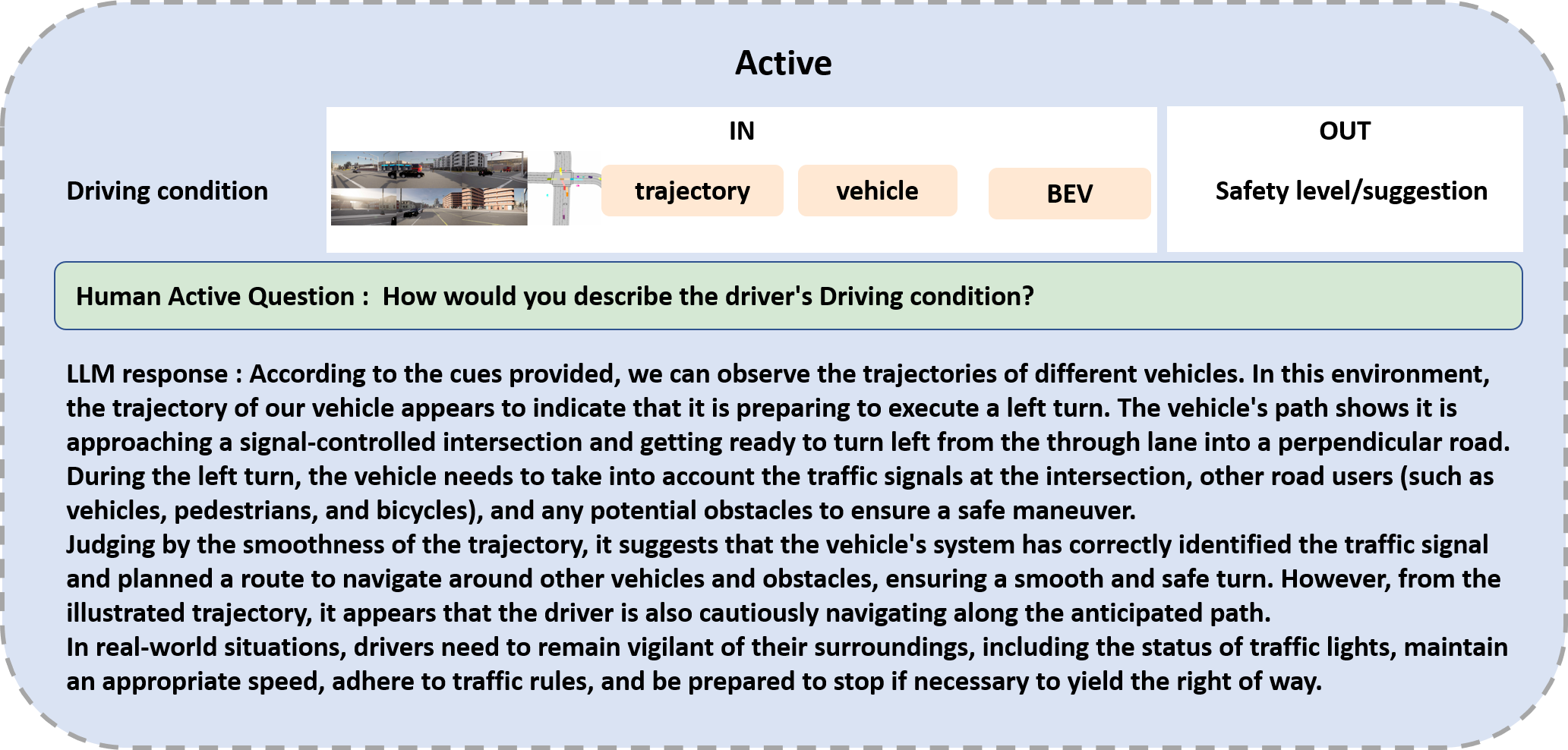}
  \caption{Sample of driving condition in AccidentGPT.}
  \label{fig:app02}
\end{figure}

\begin{figure}[h]
  \centering

\includegraphics[width=0.7\linewidth]{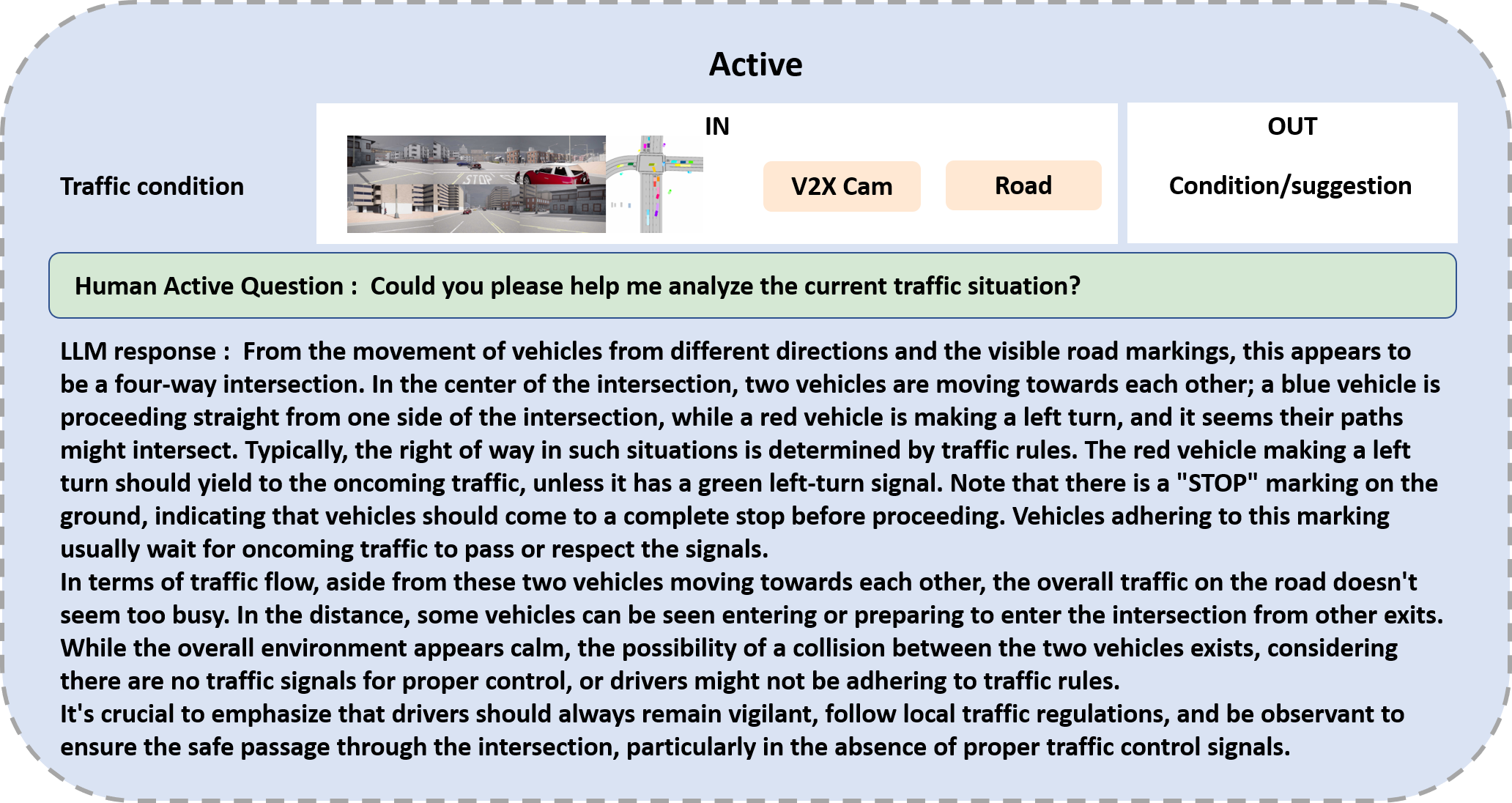}
  \caption{Sample of traffic condition in AccidentGPT.}
  \label{fig:app03}
\end{figure}

\begin{figure}[h]
  \centering

\includegraphics[width=0.7\linewidth]{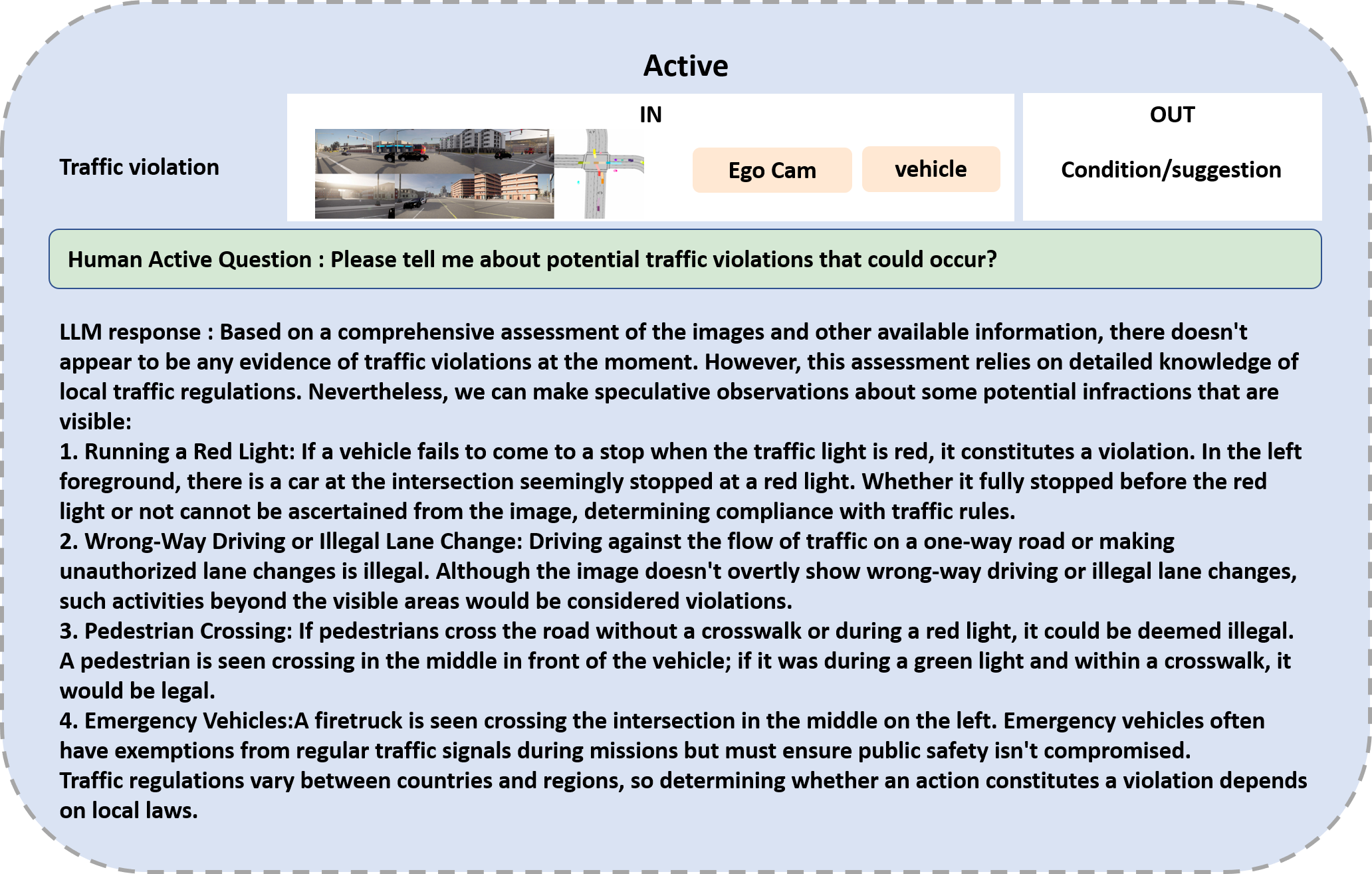}
  \caption{Sample of traffic violation in AccidentGPT.}
  \label{fig:app04}
\end{figure}

\begin{figure}[h]
  \centering

\includegraphics[width=0.7\linewidth]{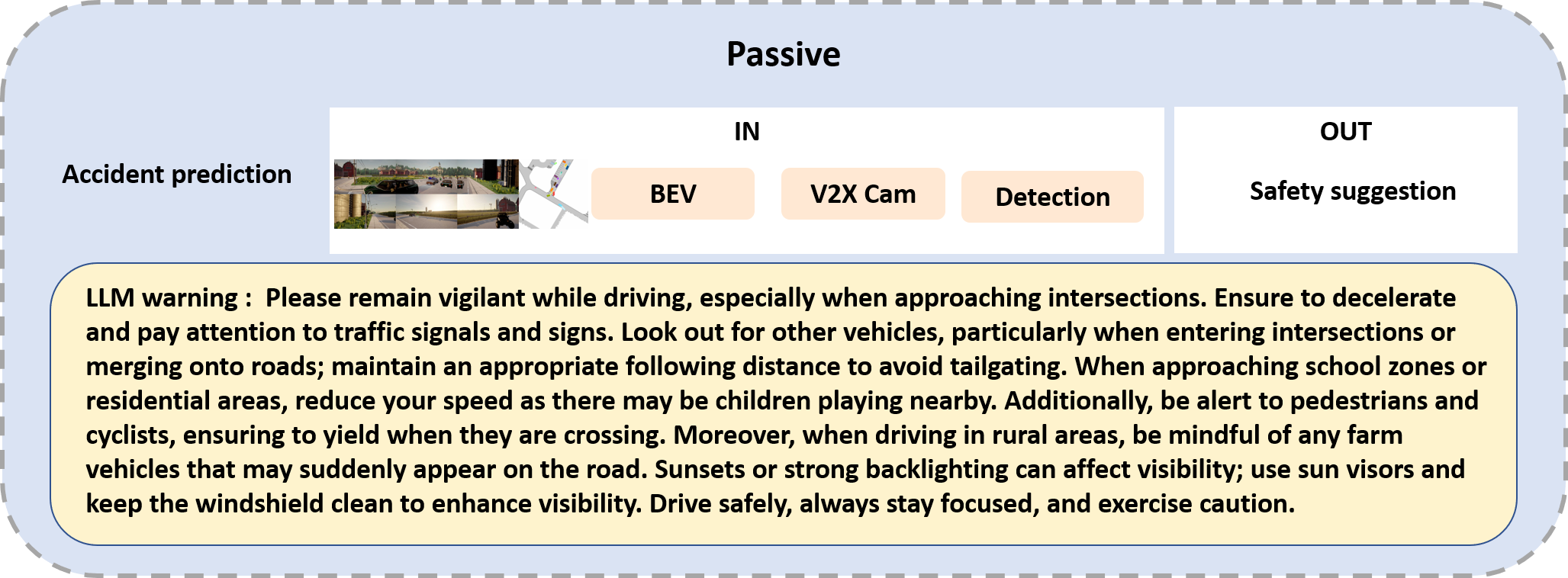}
  \caption{Sample of accident prediction in AccidentGPT.}
  \label{fig:app05}
\end{figure}

\begin{figure}[h]
  \centering

\includegraphics[width=0.7\linewidth]{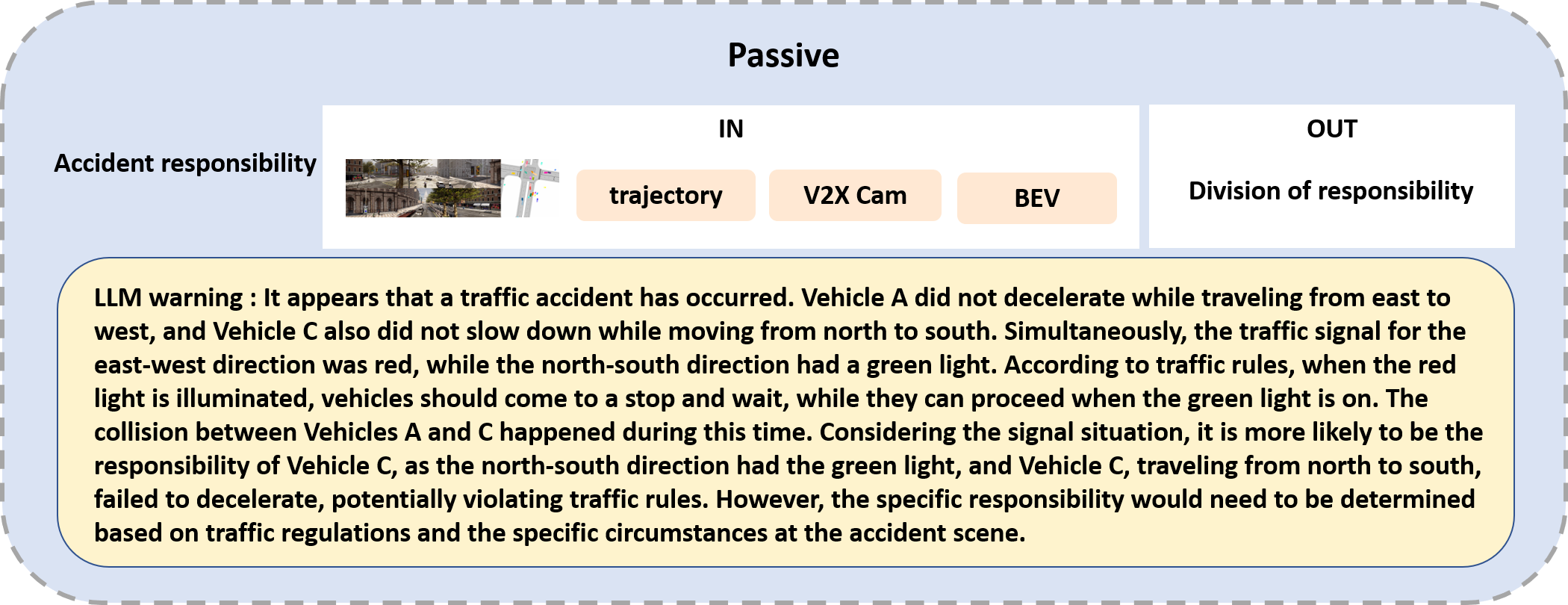}
  \caption{Sample of accident responsibility in AccidentGPT.}
  \label{fig:app06}
\end{figure}

\begin{figure}[h]
  \centering

\includegraphics[width=0.7\linewidth]{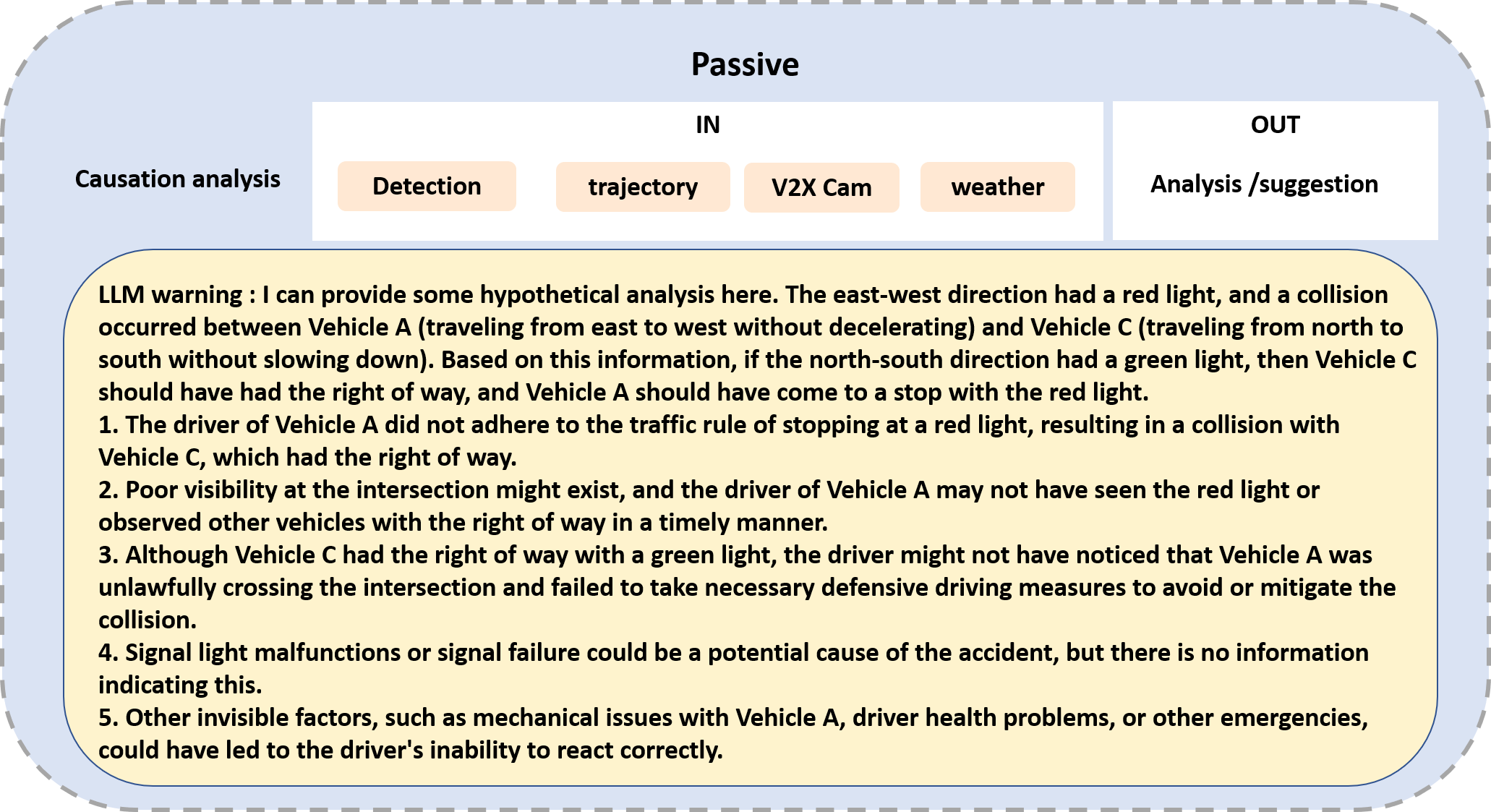}
  \caption{Sample of causation analysis in AccidentGPT.}
  \label{fig:app07}
\end{figure}

\begin{figure}[h]
  \centering

\includegraphics[width=0.7\linewidth]{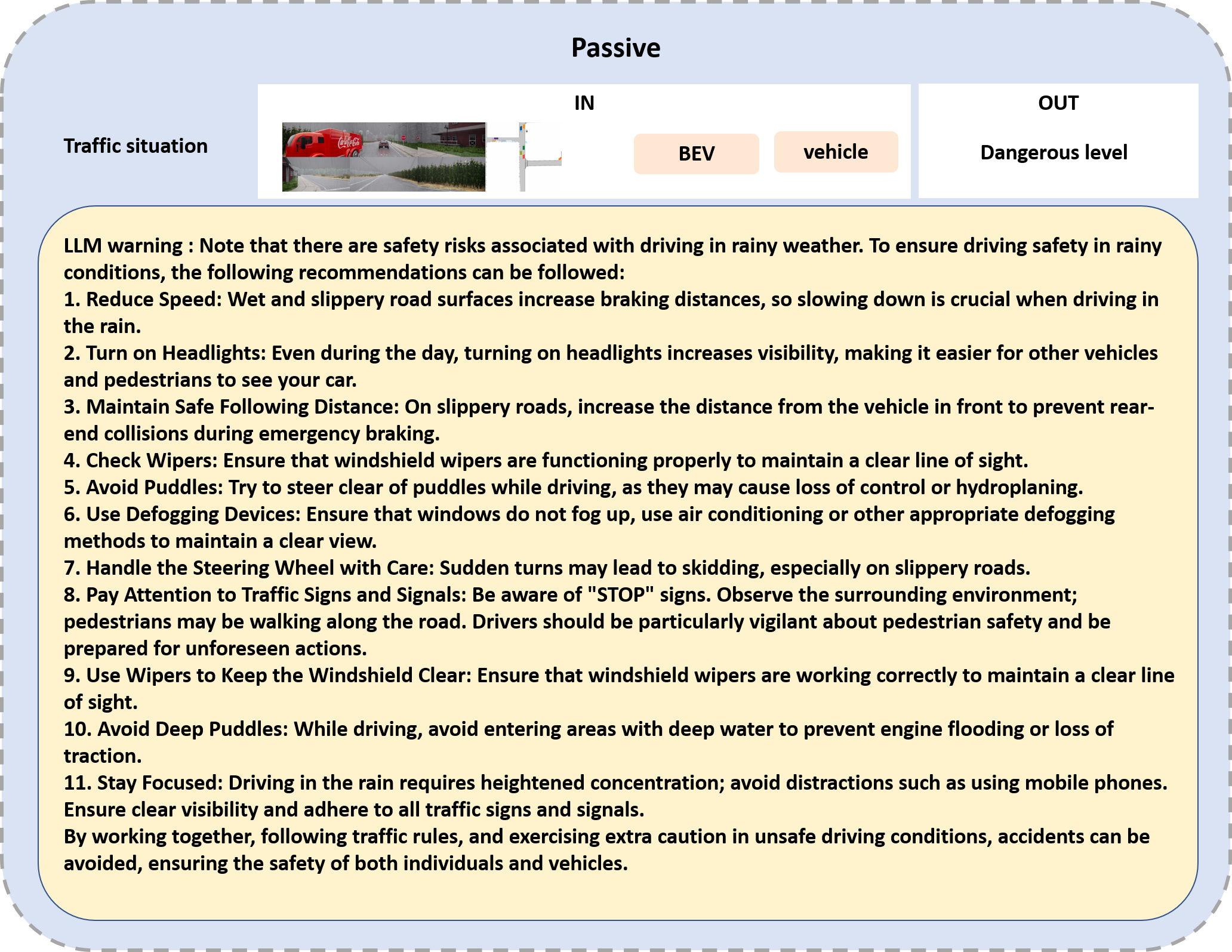}
  \caption{Sample of traffic situation in AccidentGPT.}
  \label{fig:app08}
\end{figure}

\end{document}